\documentclass[%
 preprint,
superscriptaddress,
 amsmath,amssymb,
 aps,
]{revtex4-2}

\usepackage{graphicx}
\usepackage{dcolumn}
\usepackage{bm}
\usepackage{hyperref}
\usepackage{multirow}
\usepackage{xargs}
\usepackage[normalem]{ulem} 
\usepackage{todonotes} 

\definecolor{msm_color_1}{HTML}{996600}
\definecolor{msm_color_2}{HTML}{006699}
\definecolor{msm_color_3}{HTML}{660099}
\definecolor{msm_color_4}{HTML}{990066}

\newcommandx{\unsure}[2][1=]{\todo[size=\scriptsize,linecolor=red,backgroundcolor=msm_color_1!25,bordercolor=red,#1]{#2}}
\newcommandx{\improvement}[2][1=]{\todo[size=scriptsize,linecolor=blue,backgroundcolor=msm_color_2!25,bordercolor=blue,#1]{#2}}
\newcommandx{\improve}[2][1=]{\todo[size=scriptsize,linecolor=green,backgroundcolor=msm_color_3!25,bordercolor=green,#1]{#2}}
\newcommandx{\change}[2][1=]{\todo[size=scriptsize,linecolor=green,backgroundcolor=msm_color_3!25,bordercolor=green,#1]{#2}}
\newcommandx{\suggest}[2][1=]{\todo[size=scriptsize,linecolor=black,backgroundcolor=msm_color_4!25,bordercolor=black,#1]{#2}}

\renewcommand{\a}{\mathbf{a}}

\renewcommand{\j}{\mathbf{j}}
\renewcommand{\k}{\mathbf{k}}
\newcommand{\M}[1]{\mathcal{M}_{#1}}
\newcommand{\p}{\mathbf{p}}
\renewcommand{\r}{\mathbf{r}}
\renewcommand{\u}{\mathbf{u}}
\renewcommand{\v}{\mathbf{v}}

\newcommand{\fML}[1]{f^M_{#1,L}}
\newcommand{\fMG}[1]{f^M_{#1,0}}

\newcommand{\deltaM}[1]{\delta \mathcal{M}_{#1} }

\newcommand{\tildetw}{\tilde{\tau} \tilde{\omega}}
\newcommand{\tildetwc}{\tilde{\tau} \tilde{\omega}_\tau}

\begin{document}


\title{Conservative dielectric functions and electrical conductivities from the multicomponent Bhatnagar-Gross-Krook equation}


\author{Thomas Chuna}
  \email{chunatho@msu.edu}
  \affiliation{Physics and Astronomy, Michigan State University, East Lansing, Michigan 48824, USA}%
  \homepage{https://murillogroupmsu.com/thomas-chuna/}
\author{Michael S. Murillo}%
  \email{murillom@msu.edu}
  \affiliation{Computational Mathematics, Science and Engineering, Michigan State University, East Lansing, Michigan 48824, USA}%
  \homepage{https://murillogroupmsu.com/dr-michael-murillo/}
\date{\today}

\begin{abstract}
A considerable number of semi-empirical and first-principles models have been created to describe the dynamic response of a collisionally damped charged-particle system.
However, known challenges persist for established dynamic structure factors (DSF), dielectric functions, and conductivities.
For instance, the semi-empirical Drude-Smith conductivity [N.M. Smith, Phys. Rev. B 64, 155106 (2001)] lacks interpretability, and the first-principles Mermin dielectric function [N.D. Mermin, Phys. Rev. B, 1, 2362 (1970)] does not satisfy the frequency sum rule [G.S. Atwal and N.W. Ashcroft, Phys. Rev. B 65, 115109 (2002)]. In this work, starting from the multicomponent Bhatnagar-Gross-Krook (BGK) kinetic equation, we produce a multi-species susceptibility that conserves number and momentum, which we refer to as the “completed Mermin” susceptibility, and we explore its properties and uses.
We show that the completed Mermin susceptibility satisfies the frequency sum (f-sum) rule. We compute the associated DSF and find that momentum conservation qualitatively impacts the DSF's shape for a carbon-contaminated deuterium and tritium plasma under NIF hot-spot conditions.
In the appendices, we provide numerical implementations of the completed Mermin susceptibility, for the reader’s convenience.
Further, we produce a new non-Drude conductivity model, by taking the single-species limit and introducing free parameters in the terms that enforce number and momentum conservation.
To illustrate how number and momentum conservation impact the dynamical conductivity shape, we apply our conductivity model to dynamical gold conductivity measurements [Z. Chen, et al., Nature communications, 12.1, 1638, (2021)].
Finally, comparing our model to the Drude-Smith conductivity model, we conclude that Smith’s phenomenological parameter violates local number conservation.

\end{abstract}

\maketitle

\section{Introduction \label{sec_intro}}
Many scientific investigations make use of dynamic response models, \textit{e.g.}, for the dynamic structure factor (DSF), dielectric function, and conductivity. One application is X-ray Thompson scattering (XRTS) diagnostics, which use a DSF model to infer the plasma's properties, \textit{e.g.}, number density, temperature, and mean ionization \cite{Glenzer2009}, \cite{Graziani2014}. 
Another application arises in optical conductivity experiments, which use a conductivity model to extrapolate a material's DC conductivity from optical-regime measurements \cite{Celliers2018}.
Similarly, in density functional theory (DFT) calculations of the electrical conductivity, a conductivity model is needed to extrapolate a material's DC conductivity from Kubo-Greenwood estimates \cite{Pozzo2011}.
Additionally, dielectric functions are used to estimate a plasma's stopping power \cite{ziegler2010srim}, \cite{hentschel2023improving} and DSF models are used to compute Bremsstrahlung emissions rates \cite{Ichimaru1991}. 
This work provides conservative models of the dielectric function, dynamic structure factor (DSF), and conductivity to be used in these applications and more.

These applications are addressed by a wide variety of models. We term one grouping of models ``semi-empirical''; models in this grouping have free parameters tuned to match data. Radiation hydrodynamics codes use these models because, unlike first-principles models, they can inform dynamic response over large density and temperature regimes. For example, Cochrane et al.'s work used the Lee-More-Desjairlais semi-empirical conductivity model after fitting the model to DFT estimates \cite{desjarlais2001practical}. However, semi-empirical models can lack clear interpretations. For example, non-Drude conductivity models, which are used by both high energy density (HED) and nanomaterials scientists \cite{chen2021}, \cite{lloyd2012review}, \cite{kuvzel2020review}. When non-Drude behavior is observed, the primary alternative is the Drude-Smith conductivity model \cite{Smith1968}. However, Smith’s model has a phenomenological parameter of unknown meaning \cite{Nemec2009}, \cite{shimakawa2016experiment}. Smith maintains that his modification to the Drude model includes charge carrier back scattering \cite{Smith2001}, while Cocker et al. contest Smith's claim, having derived a similar modification by assuming localized charge carriers \cite{Cocker2017}. Thus, interpretable alternatives to semi-empirical conductivity models are needed. 

We call the alternative to semi-empirical models ``first-principles'' models.
These models have calculated inputs rather than tuned parameters and, as a result, have clearer interpretations. We will consider Mermin’s collision-corrected dielectric function \cite{Mermin1970}, which requires a calculated dynamical collision frequency. To improve the Mermin model's predictions, investigators have improved collision-frequency estimates \cite{Plagemann2012}, \cite{Sperling2015}, \cite{witte2017}, \cite{hentschel2023improving}. Yet, regardless of collision-frequency choice, the Mermin model predicts nonphysical behavior. The Mermin model’s nonphysical behavior has been tied to its lack of momentum conservation. Atwal and Ashcroft showed that Mermin’s dielectric function does not have an infinitesimal plasmon width and thus does not satisfy the frequency sum (f-sum) rule \cite{Atwal2002}. Further, they show that a dielectric function that includes momentum and energy conservation has an infinitesimal plasmon width and thus does satisfy the frequency sum rule. Morawetz and Fuhrmann established that Mermin’s dielectric function scales incorrectly in the high-frequency limit, but the inclusion of momentum conservation corrects this scaling \cite{Morawetz2000b}. However, both works also demonstrate that if a \textit{single-species} dielectric function conserves momentum, then it also predicts an infinite conductivity. This implies that only multi-species dielectric functions can conserve momentum without predicting infinite conductivity. Thus, a momentum-conserving multi-species first-principles model is well motivated.

In this work, we derive our momentum-conserving multi-species first-principles model from a kinetic equation in the Bhatnagar, Gross, and Krook (BGK) approximation; this is an approach that has been used previously \cite{Morawetz2000a}, \cite{Atwal2002}. In these previous works, the distribution function exponentially decays toward a target function, which is characterized by local perturbations in the chemical potential, velocity, and temperature. These perturbations are constrained to enforce number, momentum, and energy conservation and are substituted into the linearized kinetic equation to produce dynamical response functions. In this way, Selchow and Morawetz \cite{Selchow1999} derive the Mermin's single-species dielectric function from the classical Bhatnagar, Gross, and Krook (BGK) kinetic equation \cite{BGK1954} and the classical Fokker-Plank kinetic equation. Additionally, Atwal and Ashcroft derive a number-, momentum-, and energy-conserving single-species dielectric function from the BGK kinetic equation \cite{Atwal2002}. Currently, a multi-species Mermin dielectric function exists \cite{Selchow2001DSF}, but it has not been extended to include momentum conservation. We produce the first number- and momentum-conserving multi-species susceptibility from Haack et al.’s multi-species BGK kinetic equation \cite{Haack2017} and recover, as a limit, the known multi-species Mermin susceptibility. We call our result the completed Mermin dielectric function because it satisfies the frequency sum rule.

Our multi-species completed Mermin susceptibility offers a new DSF model to inform a plasma's ionic structure. Both XRTS and bremmstrahlung emission models require estimates of the ions' DSFs \cite{Graziani2014} \cite{Ichimaru1991}. In inertial confinement fusion (ICF) experiments, hydrodynamic instabilities inject the ICF capsule's ablator material, \textit{e.g.},  carbon, into the deuterium-tritium hot-spot \cite{weber2020mixing}. The carbon contaminates trigger bremsstrahlung emission and recent efforts are establishing how to use this signature to quantify local mixing \cite{Bachmann2020} \cite{weber2020mixing} \cite{Pak2020}. Such carbon contaminants create an inherently multi-species system, which is tractable with our multi-species susceptibility. We apply the completed Mermin model to understand how conservation laws impact carbon-contaminated ion-ion DSFs.

The single-species limit of our completed Mermin model offers a new non-Drude dynamic electrical conductivity model. In experimental dynamic conductivity estimates, the measured DC conductivity can be suppressed while the measured optical conductivity is enhanced relative to the Drude model. Thus, Chen et al. recommend the use of non-Drude conductivities, \textit{e.g.}, Drude-Smith when fitting optical conductivity measurements \cite{chen2021}. We apply the completed Mermin model to understand how conservation laws impact the dynamic electrical conductivity. We show that partial number and momentum conservation can also suppress the DC conductivity and enhance optical conductivity. We compare our completed Mermin model to the established Drude-Smith model and find that Smith's parameter violates conservation laws.

The structure of the paper is as follows. In Sec.~\ref{sec_kinentics}, we introduce the multi-species BGK kinetic equation and show how to conserve local number, momentum, and energy using local variations of the chemical potential, drift velocity and temperature. In Sec.~\ref{sec_results}, we develop applications of the conserving relaxation-time approximation. First, we derive the multi-species completed Mermin susceptibility and show that it satisfies sum rules that the multi-species Mermin does not. Next, we investigate the DSF of mixtures at NIF hot-spot conditions. We quantify the impact of the light-species approximation and demonstrate that momentum conservation has a qualitative impact on the shape of the DSF. Then we use the light-species approximation to observe the impact of carbon contaminants on the light-species DSF. Finally, we derive a dynamical conductivity from our single-species completed Mermin susceptibility and demonstrate how number- and momentum-conservation parameters impact the model. We compare this new conductivity model to both the Drude and Drude-Smith models to demonstrate that Smith's phenomenological parameter violates number conservation. The details required to implement our new completed Mermin susceptibility can be found in the appendices. 

\section{Kinetics\label{sec_kinentics}}
\subsection{Describing the system}
We consider a classical system containing $N$ different charged-particle species. 
Many-body particle interactions and externally applied fields govern the dynamics of a multi-species system of charged particles.
We account for the many-body interactions with an effective one-body description for each of the $N$ species.
To this end, two terms govern the dynamics of a single particle: an effective single-particle Hamiltonian (\textit{i.e.}, a mean-field interaction) and an effective single-body inter- and intra-species collisional operator.

\subsection{Formulating the system}
The dynamics of the single-particle distribution function are governed by
\begin{align}  \label{eq_kinetic}
    \left( \partial_t + \v \cdot \nabla_\r  + \a^{(1)}_i \cdot \nabla_\v \right) f_i(\r,\v,t) = \sum_j Q_{ij}.
\end{align}
The collision operator $Q_{ij}$ denotes the two-body description of the intra-species $i=j$ collisions and inter-species $i \neq j$ collisions. 
$\a^{(1)}_i$ denotes the acceleration of species $i$ from one-body forces and is defined by
\begin{align}
    m_i \a^{(1)}_i(\k,\omega)  = - \nabla_\r U^{(1)}_i(\k,\omega) \equiv - \nabla_\r U^\mathrm{ext}_i(\k,\omega) - \nabla_\r U^\mathrm{ind}_i(\k,\omega),
\end{align}
where the external potential acting on species $i$ $U^\mathrm{ext}_i$ is inherently a one-body potential, and the induced potential acting on species $i$ $U^\mathrm{ind}_i$ is an effective one-body potential that describes the electrostatic energy. 
In Fourier space, our induced potential is given by the Hartree potential energy
\begin{align} \label{eq_HartreePotential}
    U^\mathrm{ind}_i \equiv \sum_j v_{ij}(k) \delta n_j(\k,\omega),
\end{align}
which expresses that interactions $v_{ij}(k)$ between species $i$ and $j$ cause density fluctuations $\delta n_j(\k, \omega)$ in species $j$ to affect the electrostatic potential $U^\mathrm{ind}_i$ experienced by species $i$.
This formulation facilitates species dependent ion-ion interactions which include the effects of electron screening, \textit{e.g.}, screened Coulomb potential or force matched potentials \cite{stanek2021efficacy}. In the case of a single species of electrons, we will use a Coulomb potential.

In \eqref{eq_kinetic}, we assume Haack et al.'s multi-species relaxation to equilibrium \cite{Haack2017} as our collisional operator: 
\begin{align}\label{eq_QMcBGK}
    Q_{ij} \equiv \frac{1}{\tau_{ij}} \left(\M{ij}(\r,\v,t) - f_i(\r,\v,t) \right).
\end{align}
The Maxwellian target distribution $\M{ij}$ is defined by 
\begin{align}
    \M{ij} &\equiv  \frac{g_i}{\lambda_{i,th}^3}  \left(\frac{m_i}{2 \pi T_{ij}(\r,t) }\right)^{3/2} e^{-\frac{(\varepsilon_{ij}(\r,t)  - \mu_i(\r,t) )}{T_{ij}(\r,t)} }, \label{eq_Mij}
    \\ \varepsilon_{ij} &\equiv \frac{m_i}{2}(\v-\u_{ij}(\r,t) )^2,
    \\ \frac{1}{\tau_i} &\equiv \sum_j \frac{1}{\tau_{ij}},
\end{align}
where $\lambda_{i,th}$ is the thermal de Broglie wavelength and $\lambda_i(\r,t) \equiv \frac{g_i}{\lambda_{i,th}^3} e^{\mu_i(\r,t)/T_{ij}(\r,t)}$ is the local fugacity. The target velocity $u_{ij}$ and target temperature $T_{ij}$ are defined so that the collision operator satisfies the H-theorem and number-, momentum-, and energy-conservation laws.
However, in this work, we will only use $u_{ij}$ and $T_{ij}$ as expansion parameters.

The standard interpretation of this relaxation approximation is that Vlasov dynamics govern the particle's phase-space dynamics and that for every infinitesimal time interval $dt$, a fraction $dt/\tau_{ij}$ of the particles experiences a collision event that sets their velocity distribution to the target distribution $\M{ij}$. For ion-ion collisions, Haack et al. present different choices for the ion-ion relaxation times $\tau_{ij}$ \cite{Haack2017}. These relaxation times rely on the Stanton-Murillo transport (SMT) model \cite{StantonMurillo2016} which accounts for electron screening, but suppresses the frequency dependence in $\tau_{ij}$. 

\subsection{Linearizing the multi-species BGK equation}

The simplest linearization of the kinetic equation \eqref{eq_kinetic}, assumes that $f_i(\r,\v,t)$ is at global equilibrium, but this leads to a violation of conservation laws. Instead, we follow Mermin's approach and assume that the $i^\mathrm{th}$ species' distribution function $f_i(\r,\v,t)$ and the target Maxwellian $\M{ij}(\r,\v,t)$ have small deviations from a global (0) mixture (M) equilibrium distribution
\begin{align}
    f_i(\r,\v,t) &=  \lambda_i \fMG{i}(\v) + \lambda_i \delta \fMG{i}(\r,\v,t), \label{fiexpansion}
    \\ \M{ij}(\r,\v,t) &= \lambda_i \fMG{i}(\v) + \lambda_i \deltaM{ij}(\r,\v,t), \label{Mijexpansion}
\end{align}
where the global fugacity $\lambda_i \equiv \frac{g_i}{\lambda_{i,th}^3} e^{\mu_i/T}$ has been factored out from each term, and $\fMG{i}(\v)$ is the global mixture equilibrium distribution, defined as
\begin{align} 
    \fMG{i}(\v) &\equiv \left( \frac{m_i}{2 \pi T} \right)^{3/2} \exp\left( \epsilon_i  \right), \label{eq_fMG}
    \\ \epsilon_i &\equiv \frac{m_i}{2T}(\v -\u)^2
    \\ \u &\equiv \left( \sum_i m_i n_i \u_i \right) /  \sum_i m_i n_i,
    \\  T &\equiv \left( \sum_i n_i T_i \right) / \sum_i n_i.
\end{align}
Lastly, the deviations from global equilibrium define the density fluctuations, with
\begin{align} \label{eq_densityflxtns}
    \delta n_i(\r,t) &\equiv \int d \v \: \delta \fMG{i}(\r,\v,t).
\end{align}

We insert expansions \eqref{fiexpansion} and \eqref{Mijexpansion} into \eqref{eq_kinetic}, cancel the fugacity factors, and then Fourier transform the resulting equation to arrive at the following:
\begin{align} \label{eq_kinetic-helper1}
    \left( \v \cdot \k - \omega_{\tau_i} \right) \delta \fMG{i}(\k,\v,\omega) + \sum_j \frac{i}{\tau_{ij}} \deltaM{ij}(\r,\v,t) =  m_i^{-1} \k  \cdot \nabla_\v  \fMG{i}(\v) \: U^{(1)}_i.
\end{align}
The Fourier conventions and relevant steps are described in Appendix \ref{app_fourier}. 
We have grouped the $\delta \fMG{i}$ terms, and thus, the frequency has been shifted by $\omega_{\tau_j} \equiv \omega + i/ \tau_j$.

In \eqref{eq_kinetic-helper1}, $\deltaM{ij}(\r,\v,t)$ is an unknown term.
The target Maxwellian $\M{ij}$, from \eqref{eq_Mij}, recovers the global mixture equilibrium distribution $\fMG{i}$, from \eqref{eq_fMG}, when the target velocity $\u_{ij}(\r,t)$ reduces to the bulk velocity $\u$ and the target temperature $T_{ij}(\r,t)$ reduces to the bulk temperature $T$.
This implies that, at zeroth order, $\u_{ij}(\r,t) \approx \u$ and  $T_{ij}(\r,t) \approx T$. 
Therefore, $\deltaM{ij}(\r,\v,t)$ contains the local equilibrium's local deviations in chemical potential, velocity, and temperature.
We expand $\deltaM{ij}(\r,\v,t)$ to linear order, obtaining
\begin{align}
    \deltaM{ij} &= \left(\frac{\partial \mathcal{M}_{ij}}{\partial \mu_i}\Big\vert_{\mathcal{M}_{ij}= \fMG{i}} \right) \delta \mu_i + \left(\frac{\partial \mathcal{M}_{ij}}{\partial \u_{ij}}\Big\vert_{\mathcal{M}_{ij}= \fMG{i}} \right)  \delta \u_{ij} + \left(\frac{\partial \mathcal{M}_{ij}}{\partial T_{ij}}\Big\vert_{\mathcal{M}_{ij}= \fMG{i}} \right)  \delta T_{ij}.
\end{align}
We factor out $-\cfrac{\partial \fMG{i}}{\partial \epsilon_i}$ from every term and use the chain rule to reformulate $\cfrac{\partial \fMG{i}}{\partial \epsilon_i}$ as
\begin{align}
    \cfrac{\partial \fMG{i}}{\partial \epsilon_i} = \left( \k \cdot \cfrac{\partial \fMG{i}}{ \partial \v } \right) \left( \k \cdot \cfrac{\partial \epsilon_i}{ \partial \v}\right)^{-1}.
\end{align} 
This produces our final equation for $\deltaM{ij}$:
\begin{align}
    \deltaM{ij} = - \frac{m_i^{-1} \k \cdot \nabla_\v \fMG{i} }{ \k \cdot \v} \left( \delta \mu_i + \p_i \cdot \delta \u_{ij} + \left( \frac{p_i^2}{2m_i} - \mu_i \right) \frac{\delta T_{ij}}{T_{ij}} \right). \label{dmijexpansion}
\end{align}

Inserting \eqref{dmijexpansion} into \eqref{eq_kinetic-helper1} yields
\begin{align}\label{eq_deltaf0}
    \delta \fMG{i}(\k,\v,\omega) &= \frac{m_i^{-1} \: \k \cdot \nabla_\v \fMG{i}}{\v \cdot \k - \omega_{\tau_i} } \left( U^{(1)}_i +  \frac{i}{\v \cdot \k } \sum_j \frac{1}{\tau_{ij}}  \left( \delta \mu_i +  \p_i \cdot \delta \u_{ij}
    + \left( \frac{p^2_i}{2m_i} - \mu_i \right)  \frac{\delta T_{ij}}{ T} \right) \right),
\end{align}
which indicates that the external potential, the induced potential, as well as local deviations in chemical potential, velocity, and temperature can all cause perturbations from global equilibrium. 

\subsection{Incorporating conservation laws using linear perturbations}
Previous single-species models of an electron gas violated the momentum-conservation law to account for scattering events with the other species in the system.
For instance, a gas of electrons contained in a metal will scatter with phonons and impurities causing a loss of momentum rendering conductivity finite.
Since we are accounting for all $N$ species in the system, we will include momentum conservation.

In \eqref{eq_deltaf0}, $\delta \mu_i$, $\delta \u_{ij}$, and $\delta T_{ij}$ are unknowns.
We intend to enforce conservation laws by constraining these three unknowns with three collisional invariants, \textit{i.e.}, number, momentum and energy.
Our collisional invariants are formulated as follows:
\begin{subequations}
\label{eq_multispeciesconstraints}
\begin{align}
    &\int d\v \: Q_{ij} =0, \label{eq_multispeciesmassconservation}
    \\& \int d\v \: m_i \v Q_{ij} +  \int d\v \: m_j \v Q_{ji} = 0, \label{eq_multispeciesmomentumconservation}
    \\& \int d\v \: \frac{m_i}{2} v^2 Q_{ij} +  \int d\v \: \frac{m_j}{2} v^2 Q_{ji} = 0. \label{eq_multispeciesenergyconservation}
\end{align} 
\end{subequations}
For example, satisfying \eqref{eq_multispeciesmassconservation} ensures that the same number of members of species $i$ are present before and after $i$'s collisions with species $j$. 

It may spuriously appear that our collisional invariants uniquely determine the values of local perturbations, \textit{i.e.}, $\delta\mu_i$, $\delta \u_{ij}$, and $\delta T_{ij}$.
However, notice that $\delta \mu_i$  and $\delta T_{ij}$ are accompanied by even moments of momentum in \eqref{dmijexpansion}.
Thus, the mass collisional invariant \eqref{eq_multispeciesmassconservation} and energy collisional invariant \eqref{eq_multispeciesenergyconservation} will both preserve terms with $\delta \mu_i$ and $\delta T_{ij}$, coupling these two constraints together.
Because $\delta \mu_i$ depends on a single index and $\delta T_{ij}$ depends on two indices, there remains an ambiguity about how to select $j$ for the number-density constraint  \eqref{eq_multispeciesmassconservation}. 
This issue could be resolved by changing $\delta \mu_i$ to $\delta \mu_{ij}$ and \eqref{eq_multispeciesmassconservation} to
\begin{align}
    \int d\v \: m_i Q_{ij} + \int d\v \: m_j Q_{ji} = 0.
\end{align}
However, this mass-conservation constraint would allow the system to convert species $i$ into species $j$ to reach chemical equilibrium, which is unphysical.
Therefore, no attempt is made to conserve number, momentum and energy simultaneously. 
Instead, we limit ourselves to the iso-thermal $\delta T_{ij}$ case, conserving only number and momentum.
We explore the error introduced by the iso-thermal approximation for a single-species dielectric in section \ref{subsec_conductivity}. In the single-species limit, we find that energy conservation corrections enter at order $k^2$.

We can conserve momentum even though $\delta \u_{ij}$ also depends on $i,j$, because $\delta \mu_i$  and $\delta \u_{ij}$ are accompanied by even and odd powers of momentum, respectively, in \eqref{dmijexpansion}. 
Therefore, $\delta \u_{ij}$ does not appear in the mass collisional invariant \eqref{eq_multispeciesmassconservation}, and $\delta \mu_i$ does not appear in the momentum collisional invariant \eqref{eq_multispeciesmomentumconservation}.
Hence, these equations are decoupled.
Evaluating constraints \eqref{eq_multispeciesmassconservation} and \eqref{eq_multispeciesmomentumconservation} produces 
\begin{subequations} 
\begin{align}
    \delta\mu_i &= - \delta n_i /  B^M_{i,0}, \label{eq_deltamu}
    \\ \k \cdot \delta \u_{ij} &=  \omega \left( \frac{m_i \delta n_i \tau_{ji}  + m_j \delta n_j \tau_{ij} }{ m_i n_{0,i} \tau_{ji} + m_j n_{0,j} \tau_{ij}} \right), \label{eq_deltau}
\end{align}
\end{subequations}
where
\begin{subequations} 
\begin{align}
    C^M_{i,n}(\k,\omega) &\equiv \int d\v |\p|^n \frac{m_i^{-1} \k \cdot \nabla_\v \fMG{i}}{\v \cdot \k - \omega}, \label{eq_Cn}
    \\ B^M_{i,n} &\equiv  C^M_{i,n}(\k,0). 
\end{align}
\end{subequations}
The momentum integration is carried out in Appendix \ref{app_collisionoperatormoment}.
Equations \eqref{eq_deltamu} and \eqref{eq_deltau} constrain the unknowns in our system's dynamic response.

\section{Results \label{sec_results}}
\subsection{Susceptibilities}
We turn to examining the susceptibility, from which other quantities, \textit{e.g.}, the dielectric function and the dynamic structure factor, will be produced.
The susceptibility quantifies an external potential's ability to cause density fluctuations $\delta n$ (see \eqref{eq_densityflxtns}) about the global equilibrium: 
\begin{align}\label{eq_suscept}
    \delta n_i(k,\omega) = \sum_j \chi_{ij}(k,\omega) U^\mathrm{ext}_j(k,\omega).
\end{align}
The $i$ index runs over all species ($i = 1,\dots,N$).
We formulate the system's dynamical response by integrating \eqref{eq_deltaf0} over velocity; the steps are shown in Appendix \ref{app_fractionexpansion}.
Our final result is
\begin{align} \label{eq_deltan}
    \delta n_i  &=  C^M_{i,0}(U^\mathrm{ext}_i  + \sum_j v_{ij} \delta n_j ) + \frac{i}{\omega_{\tau_i} \tau_i} \left(C^M_{i,0} - B^M_{i,0} \right) \delta \mu_i(\k) \nonumber
    \\ & \quad+ C^M_{i,0} \frac{m_i}{k^2} \sum_j \frac{i}{\tau_{ij}} \k \cdot \delta \u_{ij} + \sum_j \frac{i}{\tau_{ij}}   \left( \left( \frac{C^M_{i,2} - B^M_{i,2}}{2m_i} \right) - \mu_i ( C^M_{i,0} - B^M_{i,0} ) \right) \frac{\delta T_{ij}}{T_{ij}}.
\end{align}
The $j$ index runs over all species ($j = 1,\dots,N$).
Substituting expressions $\delta \mu_i$ from \eqref{eq_deltamu}, $\delta \u_{ij}$ from \eqref{eq_deltau}, and $\delta T_{ij}=0$ yields
\begin{align} \label{eq_deltan_1}
    \delta n_i  = C^M_{i,0}(U^\mathrm{ext}_i  + \sum_j v_{ij}(k) \delta n_j ) - \frac{i}{\omega_{\tau_i} \tau_i} \left(C^M_{i,0} - B^M_{i,0} \right) \delta n_i / B^M_{i,0}  \nonumber
    \\ + i C^M_{i,0} \frac{m_i \omega }{k^2} \sum_j \frac{1}{\tau_{ij}} \left( \frac{m_i \delta n_i \tau_{ji}  + m_j \delta n_j \tau_{ij} }{ m_i n_{0,i} \tau_{ji} + m_j n_{0,j} \tau_{ij}} \right) . 
\end{align}
Grouping the $\delta n_i$ terms yields
\begin{align}\label{eq_deltan_2}
    C^M_{i,0} U^\mathrm{ext}_i =& \delta n_i - v_{ii}(k) C^M_{i,0} \delta n_i 
    - \frac{i}{\omega_{\tau_i} \tau_i} \left( 1 - \frac{C^M_{i,0}}{B^M_{i,0}} \right) \delta n_i - \frac{ i \omega m_i}{k^2 n_i \tau_{ii} } \: C^M_{i,0} \delta n_i \nonumber
    \\ &- C^M_{i,0} \sum_{j\neq i} v_{ij}(k) \delta n_j  - \frac{i \omega m_i }{k^2} \: C^M_{i,0} \sum_{j\neq i} \frac{1}{\tau_{ij}} \left( \frac{m_i \delta n_i \tau_{ji}  + m_j \delta n_j \tau_{ij} }{ m_i n_{0,i} \tau_{ji} + m_j n_{0,j} \tau_{ij}} \right).  
\end{align}

Next, we limit ourselves to two species and solve for the susceptibility using the 2x2 matrix-inversion formula.
The result is
\begin{subequations}
\label{eq_multispeciesCM}
\begin{align}
    \chi_{11,CM} &= C^M_{1,0} \varepsilon_b^*/D,
    \\ \chi_{22,CM} &=  C^M_{2,0} \varepsilon_a^*/ D,
    \\ \chi_{12,CM} &= C^M_{1,0} \left( v_{12}(k) C^M_{2,0} + i \frac{\omega}{ k^2 } \: \frac{ m_1 m_2}{ m_1 n_{0,1} \tau_{21}+m_2 n_{0,2} \tau_{12}} C^M_{2,0} \right) /D,
    \\ \chi_{22,CM} &= C^M_{2,0} \left( v_{21}(k)  C^M_{1,0} + i \frac{\omega}{ k^2 } \frac{ m_1 m_2}{m_1 n_1 \tau_{21} + m_2 n_2 \tau_{12}}  C^M_{1,0} \right) / D,
    \\ D  &= \varepsilon_a^* \varepsilon_b^* - C^M_{1,0}\left( v_{12}(k) + i \frac{\omega}{ k^2 } \: \frac{ m_1 m_2}{m_1 n_{0,1} \tau_{21} + m_2 n_{0,2} \tau_{12}}  \right)
    \\ & \quad \quad \quad \quad \quad \quad \times C^M_{2,0} \left( v_{21}(k) + i \frac{\omega}{ k^2 } \: \frac{ m_1 m_2}{m_1 n_{0,1} \tau_{21} + m_2 n_{0,2} \tau_{12}}  \right),
    \\ \varepsilon_a^* &= 1 - v_{11} C^M_{1,0} - \frac{i}{\omega_{\tau_1} \tau_1} \left( 1 - \frac{C^M_{1,0}}{B^M_{1,0}} \right)
    - i  \frac{ \omega}{k^2} \left( \frac{m_1}{n_{1,0} \tau_{11} }
    +  \frac{ \tau_{21} }{ \tau_{12} } \frac{m_1^2}{ m_1 n_{0,1} \tau_{21} + m_2 n_{0,2} \tau_{12}} \right) C^M_{1,0},
    \\ \varepsilon_b^* &= 1 - v_{22} C^M_{2,0} - \frac{i}{\omega_{\tau_2} \tau_2} \left( 1 - \frac{C^M_{2,0}}{B^M_{2,0}} \right)
    - i  \frac{ \omega}{k^2} \left( \frac{m_2}{n_{2,0} \tau_{22} }
    +  \frac{ \tau_{12} }{ \tau_{21} } \frac{m_2^2}{ m_1 n_{0,1} \tau_{21} + m_2 n_{0,2} \tau_{12}} \right) C^M_{2,0}.
\end{align}
\end{subequations}
These equations are the primary result of this paper; we refer to them as the multi-species completed Mermin susceptibility.
For ease of use, this result is broken down into dimensionless, numerically implementable equations in Appendix \ref{app_specialfxns}.
We recover the multi-species Mermin-like susceptibility \cite{Ropke1999, selchow2002extended} from \eqref{eq_multispeciesCM} by neglecting the terms with a factor of $\omega/k^2$.
If we also let the relaxation times $\tau_{ij}$ go to infinity, we recover the random-phase approximation (RPA).
Notice that in both the Mermin and the RPA susceptibilities, the interaction potential $v_{12}(k)$ is the only term coupling species $1$'s susceptibility to species $2$'s susceptibility.
Whereas, in the completed Mermin susceptibility \eqref{eq_multispeciesCM}, the conservation of momentum also couples the susceptibilities. 

We recover the single-species limit of the completed Mermin model by neglecting the coupling terms in \eqref{eq_multispeciesCM},
\begin{align} \label{eq_singlespeciesCM}
    \chi^{CM} \equiv \cfrac{C_0}{ \varepsilon - \cfrac{i}{\omega_{\tau} \tau} \left( 1 - \cfrac{C_0}{B_0} \right)  - \frac{i m \omega}{ k^2 n_0 \tau}C_0 },
\end{align}
whereas the single species Mermin is given by
\begin{align} \label{eq_singlespeciesM}
    \chi^M \equiv \cfrac{C_0}{ \varepsilon - \cfrac{i}{\omega_{\tau} \tau} \left( 1 - \cfrac{C_0}{B_0} \right) }.
\end{align}
In both expressions we neglect indices since there is only one species. Comparing \eqref{eq_singlespeciesCM} with \eqref{eq_singlespeciesM}, the completed Mermin susceptibility includes the momentum-conservation correction
\begin{align} \label{eq_singlespeciesmomentumconservation}
    \frac{i m \omega}{ k^2 n \tau} \: C_0.
\end{align}
The form of this momentum-conservation correction \eqref{eq_singlespeciesmomentumconservation} matches Morawetz and Fuhrmann's single-species local field correction \cite{Morawetz2000b}.
The correction \eqref{eq_singlespeciesmomentumconservation} arises from the single-species version of the momentum constraint \eqref{eq_deltau}, with
\begin{align}
    \k \cdot n_0 \delta \u = \omega \delta n.
\end{align} 
Comparatively, Mermin produced his number-conservation constraint by enforcing $\k \cdot \j = \omega \delta n$ \cite{Mermin1970}.
Our momentum constraint, \eqref{eq_deltau}, differs only in that $\j = n_0 \: \delta \u$.
This suggests that enforcing $\j = n_0 \: \delta \u$ in the Mermin continuity equation and varying local equilibrium with respect to velocity leads to momentum conservation.
Thus, we refer to our susceptibility as the ``completed Mermin'' susceptibility.

\subsection{Dielectric functions}
The dielectric function is defined as
\begin{align} \label{eq_dielectric}
    \frac{1}{\varepsilon(k,\omega)} = 1 + \sum_{ij} v_{ij}(k) \chi_{ij}(k,\omega).
\end{align}
An essential property of every dielectric function is the fulfillment of sum rules, which determines the quality of the dielectric function moment by moment \cite{boonyip1991}.
The frequency sum (f-sum) rule expresses whether the local continuity equation is satisfied; it is expressed as
\begin{align}  \label{eq_fsumrule-eps}
    \int_{-\infty}^{\infty} d\omega \: \omega \text{Im} \{ \varepsilon^{-1} \} = - \pi \sum_i \frac{\omega_{p,e}^2}{k_{D,e}^2} \frac{m_e}{m_i} \frac{n_i}{T} v_{ii}(k) k^2.
\end{align}
For an isolated species of electrons in the unscreened limit, the RHS of \eqref{eq_fsumrule-eps} reduces to the familiar $- \pi \omega_p^2$. Another sum rule is the perfect-screening sum rule, which is valid when there is no appreciable $k$ dependence, that is, when the relation between the induced density and the external potential is a purely local one \cite{hansen2013theory}; it is expressed as
\begin{align} \label{eq_screeningsumrule-eps}
    \underset{k \rightarrow 0}{\lim} \: \: \underset{\tilde{\lambda}_s^{-1} \rightarrow 0}{\lim}\: \int_{-\infty}^{\infty} \frac{d\omega}{\omega} \text{Im} \{ \varepsilon^{-1} \} = - \pi .
\end{align}
For the screening sum rule, the no-screening limit (\textit{i.e.}, $\tilde{\lambda}_s^{-1} \rightarrow 0$) is taken first, and the long-wavelength limit (\textit{i.e.}, $k \rightarrow 0$) is taken second.

For the one-component plasma (OCP) of electrons, we produce analytic expressions of $\underset{\k \rightarrow 0}{\lim} \: \text{Im}\{ \varepsilon^{-1} \}$ for each known collisional single-species case: Mermin (M) \eqref{eq_singlespeciesM}, our completed Mermin (CM) \eqref{eq_singlespeciesCM}, and Atwal-Ashcroft (AA), which locally conserves number, momentum, and energy \cite{Atwal2002}.
In terms of dimensionless parameters, the functional forms are as follows:
\begin{subequations}
\label{eq_longwavelengthexpansion}
\begin{align}
    &\underset{\k \rightarrow 0}{\lim} \: \text{Im}\left\{\frac{1}{\varepsilon^\textrm{M}} \right\} = - \frac{(\tilde{\omega}/\tilde{\tau})}{ (\tilde{\omega}/\tilde{\tau})^2 + ( \tilde{\omega}^2 - 1)^2},
    \\& \underset{\k \rightarrow 0}{\lim} \: \text{Im}\left\{\frac{1}{\varepsilon^\textrm{CM}} \right\} = - \frac{ \tilde{k}^2 \tilde{\tau} \tilde{\omega} }{ - 2 \tilde{k}^2 (\tilde{\omega}^2 - 1) + (1 + \tilde{\tau}^2 \tilde{\omega}^2) ( \tilde{\omega}^2 - 1)^2},
    \\& \underset{\k \rightarrow 0}{\lim} \: \text{Im}\left\{\frac{1}{\varepsilon^\textrm{AA}} \right\} = - \cfrac{ \tilde{k}^2 \tilde{\tau} \tilde{\omega}}{- 2 \tilde{k}^2  \left( \frac{1 + 3 \tilde{\tau}^2 \tilde{\omega}^2 - 2 \zeta^{-2} \tilde{\tau}^2}{ 1 + \tilde{\tau}^2 \tilde{\omega}^2} \right)( \tilde{\omega}^2 -1 ) + \frac{1}{3} (1 + \tilde{\tau}^2 \tilde{\omega}^2) ( \tilde{\omega}^2 -1 )^2}.
\end{align}
\end{subequations}
We have used electronic quantities to render the functions dimensionless: $\tilde{\tau} = \tau \omega_{p,e}$, $\tilde{\omega} = \omega/\omega_{p,e}$, $\tilde{k} = k/k_{D,e}$. The steps taken to compute these dimensionless expressions are shown in Appendix \ref{app_expansions}.

\begin{figure}
    \includegraphics[width=.5\textwidth]{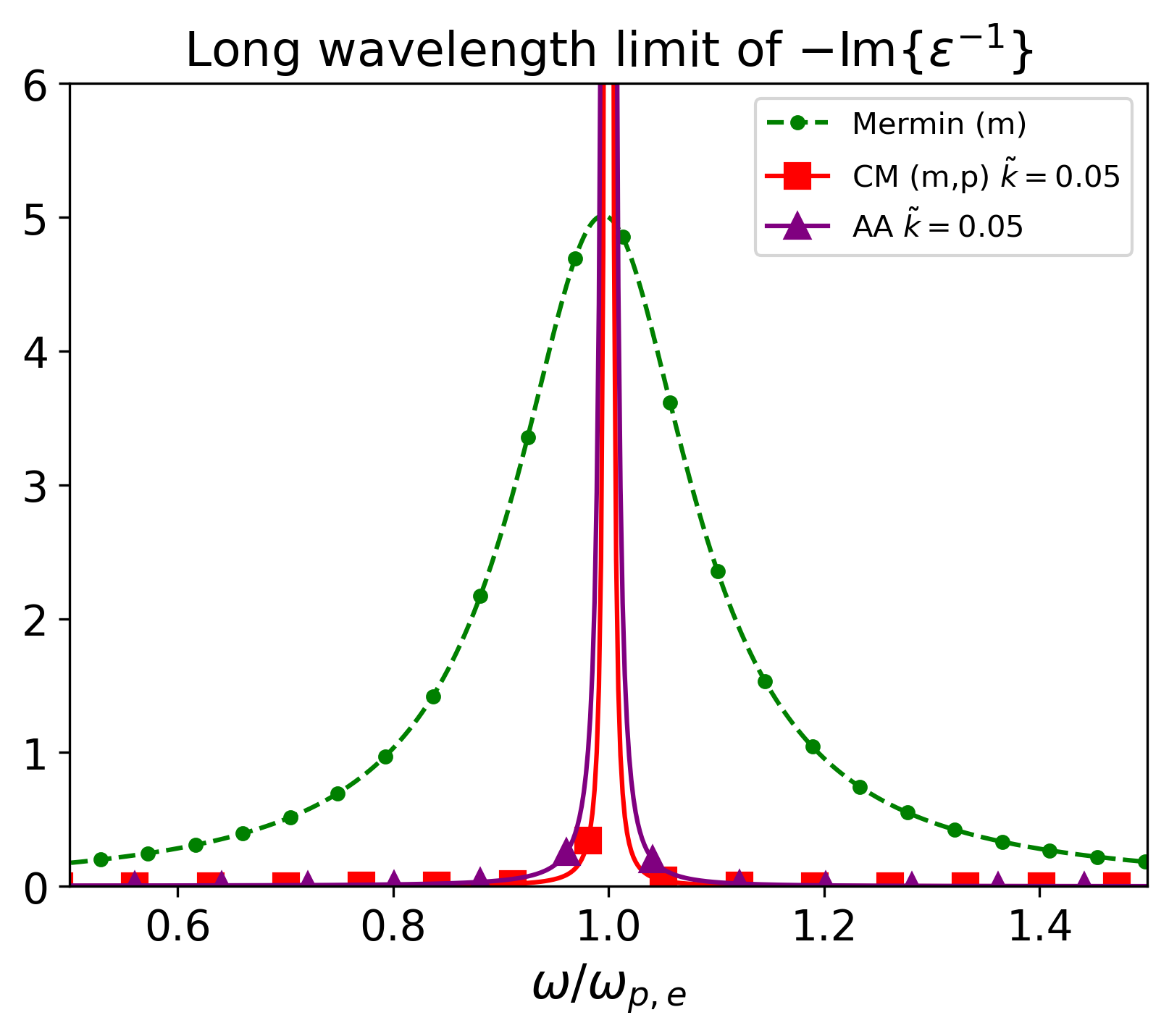}
    \caption{Plots of the long-wavelength expansion of Im$\{\varepsilon^{-1}\}$ in the one-component plasma (OCP) case, for the Mermin (green), completed Mermin (CM, red), and Atwal-Ashcroft (AA, purple) dielectric functions from \eqref{eq_longwavelengthexpansion}. We evaluate the functions at $\tilde{k} \approx 0.05$, $(\omega_{p,e} \tau)^{-1} =.2$, $\omega_{p,e} = 1 $.} 
    \label{fig:epsinv_q0lim-single}
\end{figure}

All three functional forms in \eqref{eq_longwavelengthexpansion} are Lorentzian.
Mermin's width parameter is $1/\tilde{\tau}$, while both the completed Mermin and the Atwal-Ashcroft width parameters are $\tilde{k}$. As $k \rightarrow 0$, the completed Mermin and Atwal-Ashcroft dielectric functions become Dirac deltas. We plot the long-wavelength expansions from \eqref{eq_longwavelengthexpansion} for wave number $\tilde{k} \approx 0.05$ in Fig.~\ref{fig:epsinv_q0lim-single}.
The completed Mermin and Atwal-Ashcroft functions have narrower widths than the Mermin because Mermin's width parameter $1/\tilde{\tau}$ remains finite in the long-wavelength limit.

When the width becomes infinitesimally small, we recover a Dirac delta function:
\begin{align}\label{eq_dirac}
    \text{Im}\{\varepsilon^{-1}\} = - \pi \omega \: \delta(\omega - \omega_p).
\end{align} 
By substitution, we see that the Dirac delta satisfies both sum rules \eqref{eq_fsumrule-eps} and \eqref{eq_screeningsumrule-eps}. Therefore, the completed Mermin and the Atwal-Ashcroft dielectric functions satisfy both the f-sum rule and the screening-sum rule. Because the Mermin dielectric function does not converge to a Dirac delta, it will not satisfy the f-sum rule; however, the Mermin dielectric function will satisfy the screening-sum rule because it has no $k$ dependence. In the RPA limit (\textit{i.e.}, $\tau \rightarrow \infty$), the Mermin model also becomes a Dirac delta function and satisfies the f-sum rule.

\begin{figure}[t]
    \centering
    \includegraphics[width=.46\textwidth]{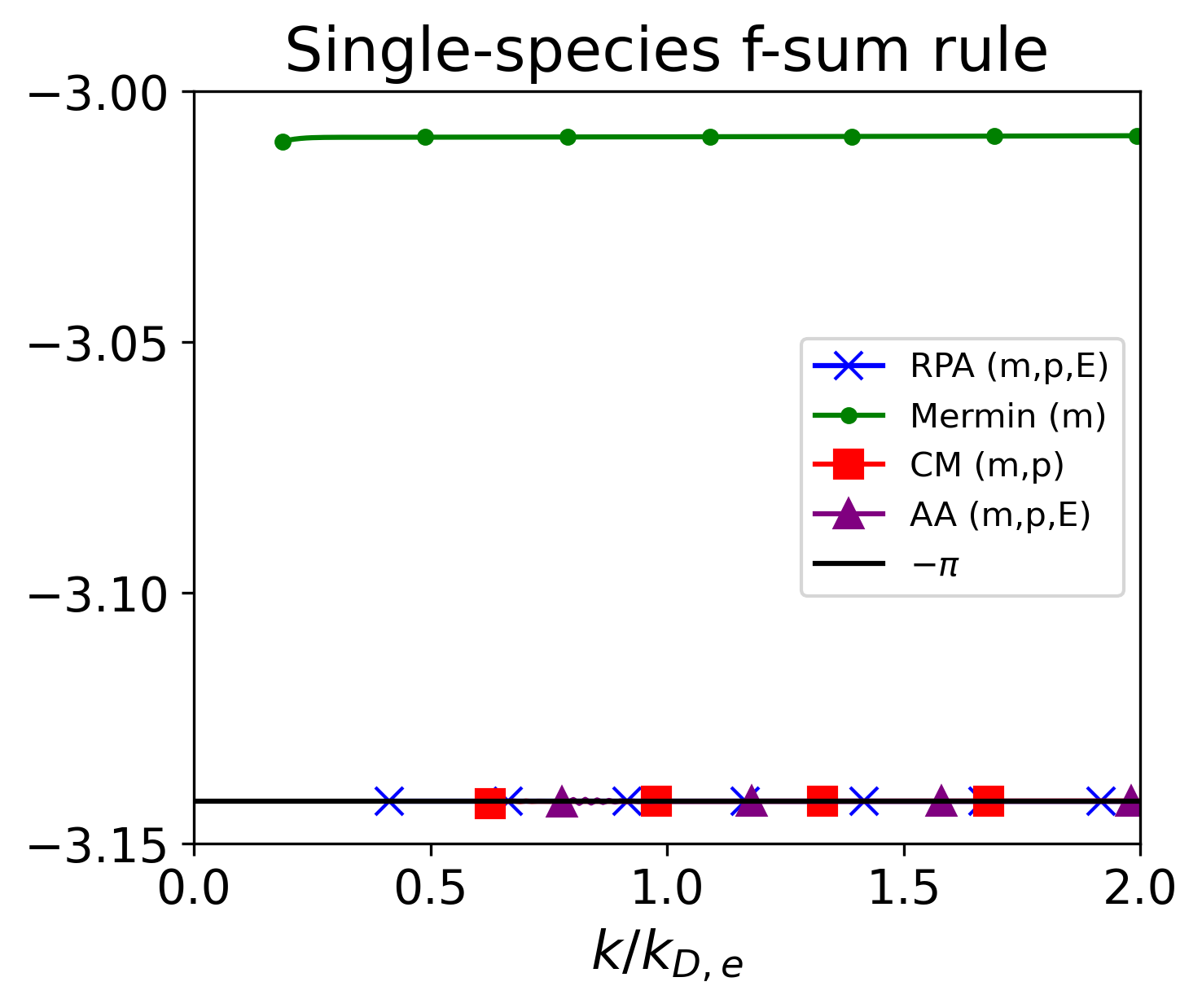}%
    \includegraphics[width=.45\textwidth]{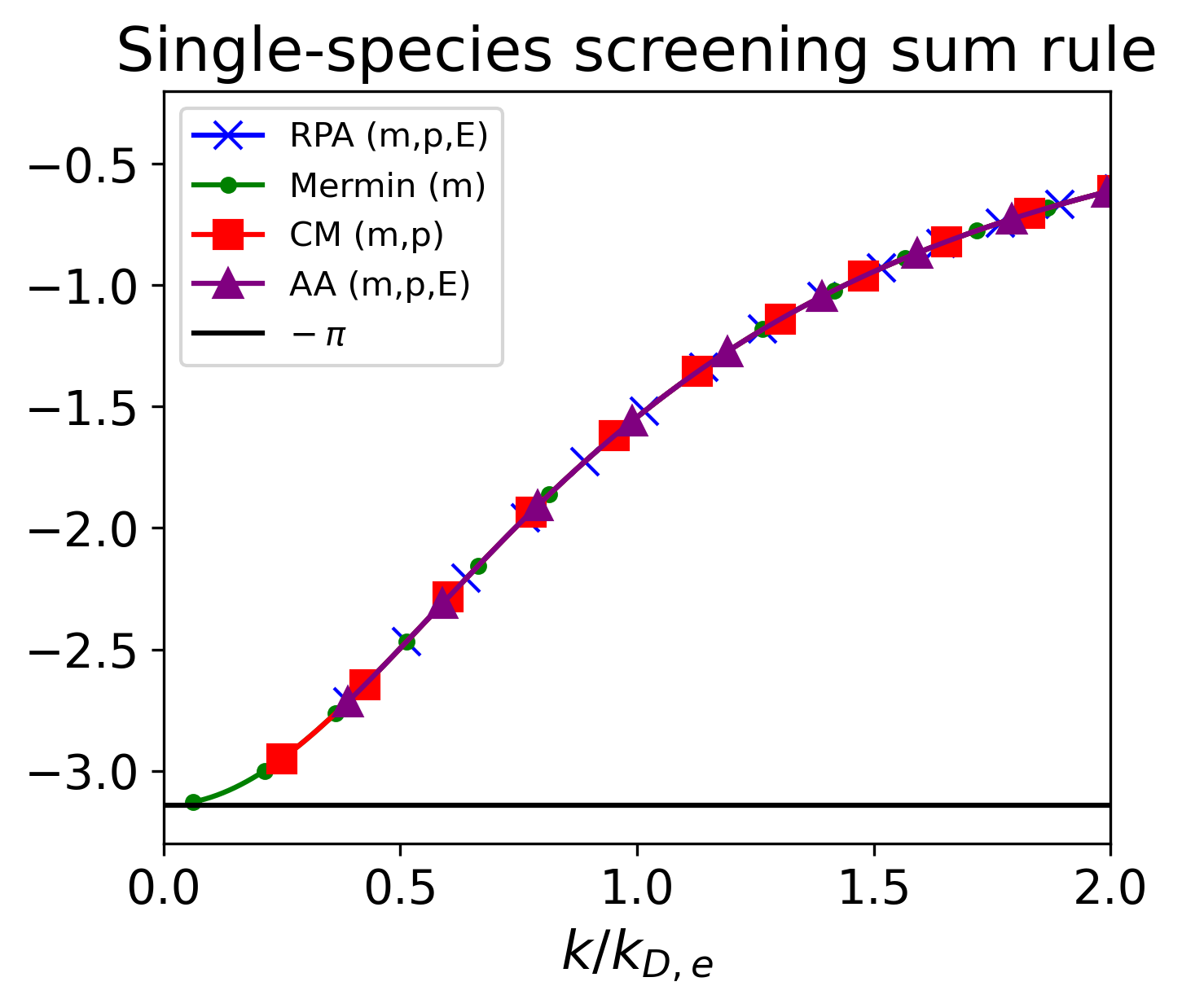}%
    \caption{Left: A plot of the frequency-sum (f-sum) rule, \eqref{eq_fsumrule-eps}, for the random-phase approximation (RPA, blue), Mermin (green), completed Mermin (CM, red), and Atwal-Ashcroft (AA, purple) dielectric functions of the Yukawa one component plasma (YOCP).
    Right: A plot of the screening-sum rule, \eqref{eq_screeningsumrule-eps}, for the single-species RPA (blue), Mermin (green), and completed Mermin (red) dielectric functions of the Yukawa one component plasma (YOCP). 
    In both plots, wiggles arise in the long-wavelength limit because the susceptibilities become Dirac deltas and numerical integration becomes impossible; we have truncated our plots before that happens.}
    \label{fig_sumrules}
\end{figure}

\begin{figure}[h]
    \centering
    \includegraphics[width=.46\textwidth]{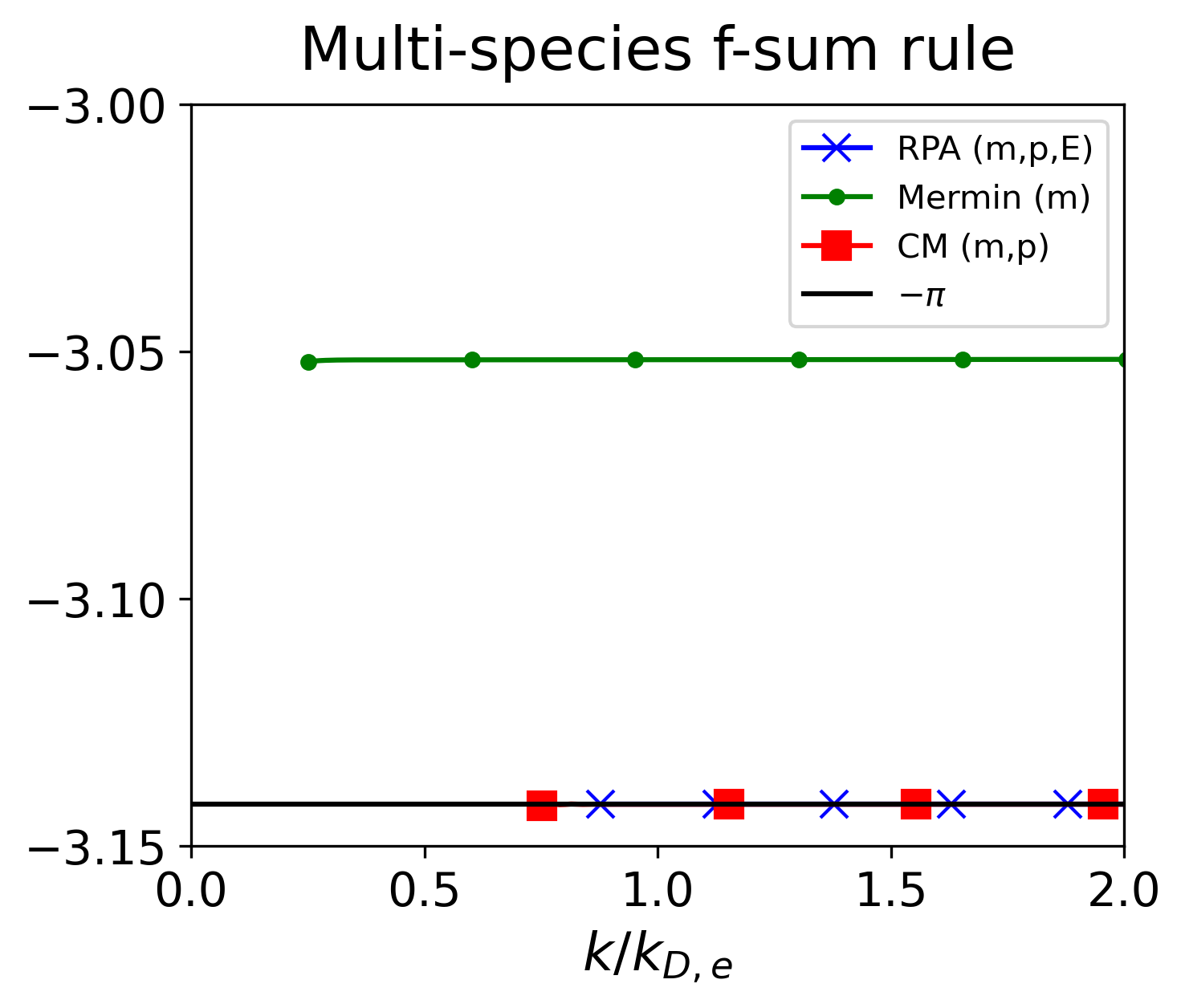}%
    \includegraphics[width=.45\textwidth]{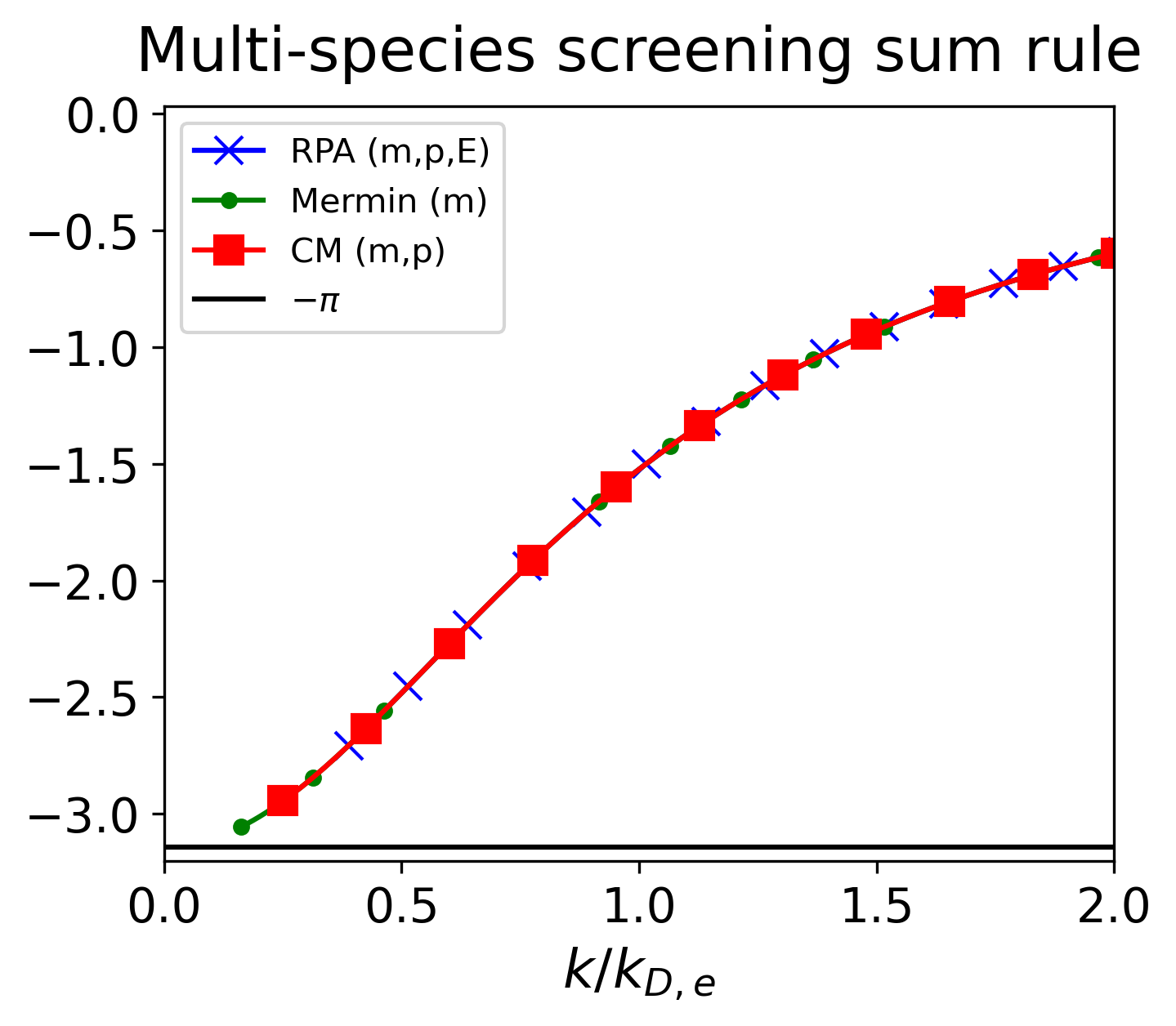}
    \caption{Left: A plot of the dielectric function frequency-sum (f-sum) rule, \eqref{eq_fsumrule-eps}, for the random-phase approximation (RPA, blue), Mermin (green), and completed Mermin (CM, red) of the binary Yukawa mixture.
    Right: A plot of the screening sum rule, \eqref{eq_screeningsumrule-eps}, for multi-species RPA (blue), Mermin (green), and completed Mermin (red) of the binary Yukawa mixture.
    In both plots wiggles arise in the long wavelength limit because the susceptibilities become Dirac deltas and numerical integration becomes impossible, we have truncated our plots before that happens.}
    \label{fig_multi-sumrules}
\end{figure}

We can also assess whether the sum rules hold outside the OCP long-wavelength limit. The sum rules are computed numerically at a finite $k$, for the Yukawa one component plasma (YOCP) and the results are shown in Fig.~\ref{fig_sumrules}. This plot shows that, unlike the Mermin dielectric function, the completed Mermin model integrates to $-\pi$ and therefore satisfies the f-sum rule. This matches the behavior seen in the long-wavelength limit of the OCP. For the screening-sum rule, all results converge to $-\pi$ in the long wavelength limit.
Additionally, Fig.~\ref{fig_multi-sumrules} shows that for the binary Yukawa mixture (BYM) the multi-species dielectric functions exhibit the same sum rule behavior as their YOCP counterparts.


\subsection{Dynamic structure factors\label{sec_ICF}}
The DSF is an essential input into HED experimental diagnostics, often determining the quality of the experimental inferences. The classical fluctuation dissipation theorem (FDT) defines the DSF as
\begin{align}
    S_{ij}(k,\omega) = \frac{-2 T}{n_i \omega} \chi_{ij}''(k,\omega),
\end{align}
where $\chi_{ij}''$ is the imaginary part of the susceptibility. 

We examine the effects of high-Z contaminants on deuterium's (D) and tritium's (T) DSFs in an inertial confinement fusion (ICF) hot-spot. 
Based on Hu et al's work on National Ignition Facility (NIF) direct drive, we assume that the hot-spot is at a mass density of $1002$ g/cc and temperature of $928$ eV \cite{Hu2010}.
We consider two BYMs.
In the first mixture, we neglect carbon (C) contaminants and consider an uncontaminated D and T plasma; we explore the impact of the light-species approximation, whereby D and T are treated as a single species.
In the second mixture, we include C contaminants in the hot-spot and use the light-species approximation to reduce the contaminated D, T, and C plasma to a mixture of only light species and C.
Including electrons as a third species would reduce to the two species case, due the mass scale separation \cite[Section 10.9]{hansen2013theory}.

\begin{figure}
    \centering
    \includegraphics[width=.9\textwidth]{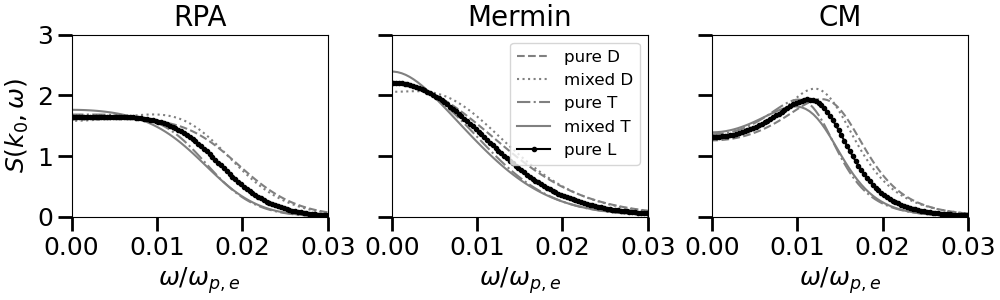}
    \caption{Plots compare $S(k,\omega)$ at fixed $k_0 = .63 k_{D,e}$ across different models of the binary Yukawa mixture DSF. Each plot contains a pure deuterium (pure D), a deuterium mixed with tritium (mixed D), a pure tritium (pure T) and a tritium mixed with deuterium (mixed T), as well as a pure light species (pure L) [defined \eqref{eq_lightapprox}]. The parameters associated to each plasma are given in Table \ref{tab_lightparams}.}
    \label{fig_multi-DSFkslice}
\end{figure}

First, neglect C contaminants.
Given mass density $\rho$ the number density of the individual components is determined by
\begin{align} \label{eq_mixturedensity}
    \rho = m_D n_D + m_T n_T.
\end{align}
We consider three different combinations of $n_T$ and $n_D$: pure D (\textit{i.e.}, $n_T=0$), pure T (\textit{i.e.}, $n_D=0$), and equal parts D and T (\textit{i.e.}, $n_D=n_T$)
Additionally, we consider the light species approximation and treat D and T as indistinguishable components of a single species plasma.
The relative number density of the tritium and deuterium sets the mass of this light (L) species.
This is formulated
\begin{subequations}
\label{eq_lightapprox}
\begin{align}
    m_L &= \frac{n_D m_D +n_T m_T}{n_D +n_T},
    \\ n_L &= n_D + n_T,
\end{align}
\end{subequations}
which implies that $\rho = n_L m_L$.
From \eqref{eq_mixturedensity} and \eqref{eq_lightapprox}, we compute the number density and the other relevant plasma parameters,  which we list in Table \ref{tab_lightparams}.
\begin{table}
\caption{\label{tab_lightparams}%
Tabulated plasma parameters for a pure deuterium (D) plasma, a pure tritium (T) plasma, a pure L plasma, and a mixed D and T plasma; all plasmas are at a mass density of $1002$ g/cc and temperature of $928$ eV  \cite{Hu2010}.
The effective charge of the ion $Z^*$ is computed using More's Thomas-Fermi ionization estimate \cite{more1985pressure}.
The coupling parameter is defined $\Gamma_i \equiv \left(Z_i^*\right)^2 / a_i T$ where $a_i = (3 Z_i^* / 4 \pi n_e)^{1/3}$ and $n_e= \sum_i Z^*_i n_i$.
The screening length is defined $\tilde{\lambda}_s \equiv (\lambda_s k_{D,e})^{-1}$, where $\lambda_s$ is the Stanton and Murillo screening length \cite{StantonMurillo2016}.
Lastly, for $\nu$ we use the temperature relaxation collision rates defined in Haack et al. \cite{Haack2017}.
}
Uncontaminated deuterium tritium plasmas
\vspace{.1cm}
\begin{ruledtabular}
\begin{tabular}{c| c c c c c c c}
           & species & n (1/cc)& $Z^*$ & $\Gamma$ & $\tilde{\lambda}^{-1}$ & $\nu_{i=j}$ (1/s) & $\nu_{i \neq j}$ (1/s)
\\ \hline 
No mix & pure L $\:$ & 2.40e26 & 0.966 & 0.147 & 1.176 & 6.30e-03 & N/A
\\ \hline 
No mix & pure D $\:$ & 2.00e26 & 0.966 & 0.138 & 1.198 & 5.42e-03 & N/A
\\ \hline 
No mix & pure T $\:$ & 3.00e26 & 0.966 & 0.158 & 1.146 & 7.56e-03 & N/A
\\ \hline
\multirow{2}{*}{$n_T = n_D$} & mixed D & 1.19e+26 & 0.965 & 0.116 & 1.196 & 3.47e-03 & 3.04e-03
\\                           & mixed T & 1.19e+26 & 0.965 & 0.116 & 1.196 & 2.84e-03 & 3.04e-03
\end{tabular}
\end{ruledtabular}
\end{table}
For the various plasmas, we plot the DSF at a fixed $k$ in Fig.~\ref{fig_multi-DSFkslice}. For a fair comparison the mixed cases plot a combined DSF (\textit{e.g.}, ``mixed D'' plots $S = S_{DD} + S_{DT}$), which indicates the correlation from a deuterium particle to any other particle. The single light species DSF is in qualitative agreement with all cases. This suggests that the light species approximation is reasonable for the DSF of a D and T plasma at $1002$ g/cc and temperature of $928$ eV. Together these plots show that all five cases have qualitative agreement within a given set of conservation laws. Thus, the light species approximation works in a given model. Furthermore, the models themselves have qualitative disagreements. The inclusion of momentum conserving collisions within the completed Mermin model produces a plasmon peak than is not present in either the Mermin or RPA models.

\begin{figure}
    \centering
    \includegraphics[width=.9\textwidth]{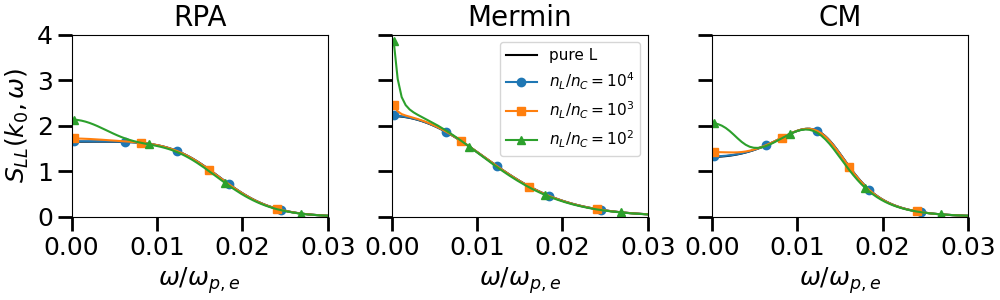}
    \caption{Plots compare carbon contaminated $S_{LL}(k,\omega)$ at fixed $k_0 = .63 k_{D,e}$ across different models of the binary Yukawa mixture DSF. Each plot repeats the pure light species plasma (pure L) [defined \eqref{eq_lightapprox}] from Fig. \ref{fig_multi-DSFkslice} and includes plasmas with increasingly greater proportion of carbon. The plasma parameters for the plasmas which contain carbon are given in Table \ref{tab_contaminatedparams}. In the plots, the least contaminated plasma ($n_L/n_C=10^4$, blue) is indistinguishable from the pure L DSF.}
    \label{fig_impurities_kappa}
\end{figure}

In the second case, we use the light species approximation for D and T and introduce C; this is formulated as 
\begin{align}
    \rho = m_L n_L + m_C n_C.
\end{align}
We assume equal amounts of deuterium and tritium and consider a 1 part carbon per $\eta$ parts light (\textit{i.e.}, $n_C=n_L/\eta$) which allow us to express the density as 
\begin{align}
    \rho &= (m_L + m_C/\eta ) n_L.
\end{align}
From this expression, we compute the number density and the other relevant contaminated plasma parameters, which we list in Table \ref{tab_contaminatedparams}.
\begin{table}
\caption{\label{tab_contaminatedparams}%
Tabulated plasma parameters for contaminated light species plasmas at three different levels of contamination: 1 carbon atom per $\eta = 10^4, 10^3, 10^2$ light species atoms.
Tabulated plasma parameters are computed in the same way as in Table.~\ref{tab_lightparams}
}
Carbon contaminated light species plasma
\vspace{.1cm}
\begin{ruledtabular}
\begin{tabular}{c| c c c c c c c}
           $\eta$ & species & n (1/cc)& $Z^*$ & $\Gamma$ & $\tilde{\lambda}_s^{-1}$ & $\nu_{i=j}$ (1/s) & $\nu_{i \neq j}$ (1/s)
\\ \hline 
\multirow{2}{*}{$10^4$} & L & 2.40e+26 & 0.965 & 0.147 & 1.176 & 6.29e-03 & 3.57e-06
\\                      & C & 2.40e+22 & 5.895 & 0.139 & 1.176 & 3.57e-02 & 2.44e-05
\\ \hline
\multirow{2}{*}{$10^3$} & L & 2.39e+26 & 0.965 & 0.146 & 1.182 & 6.25e-03 & 3.46e-05
\\                      & C & 2.39e+23 & 5.788 & 0.289 & 1.182 & 3.46e-02 & 2.33e-04
\\ \hline 
\multirow{2}{*}{$10^2$} & L & 2.29e+26 & 0.965 & 0.144 & 1.207 & 5.90e-03 & 3.10e-04
\\                      & C & 2.29e+24 & 5.584 & 0.579 & 1.207 & 3.10e-02 & 2.02e-0
\end{tabular}
\end{ruledtabular}
\end{table}
Plots indicating how carbon contaminates affect the DT DSFs for RPA, Mermin, and CM models are given in Fig.~\ref{fig_impurities_kappa}.
The pure light species DSF $S_{LL}$ is indicated by a black line and repeats the pure L from Fig.~\ref{fig_multi-DSFkslice}. The colored lines indicate increasingly greater proportion of carbon. While all models indicate that the introduction of carbon increases charge screening, the models which satisfy the f-sum rule have a qualitatively different behavior than the Mermin model.

\subsection{Conductivities \label{subsec_conductivity}}

Both optical conductivity experiments and Kubo-Greenwood conductivity estimates need a dynamical model to estimate the DC conductivity.
When the Drude conductivity model fails to fit the conductivity estimates, there are few other conductivity models to use.
To produce a new model, we apply the single species limit of the completed Mermin model to an OCP of electrons 
The dynamical conductivity can be related to the single species dielectric function models via
\begin{align} \label{eq_sigma-dielectric}
    \sigma(k,\omega) =   \frac{i \omega}{4\pi} \left( 1- \varepsilon(k,\omega) \right).
\end{align}
where $\varepsilon(k,\omega)$ is the single species limit of \eqref{eq_dielectric} and $v_{11}(k)$ is a Coulomb interaction between electrons.

Equation \eqref{eq_sigma-dielectric} requires an estimate of the dielectric function $\varepsilon(k,\omega)$.
Using \eqref{eq_singlespeciesCM} and \eqref{eq_dielectric}, the long wavelength expansion of the completed Mermin dielectric function, with a Coulomb interaction, yields
\begin{align}\label{eq_epsAAk0}
    \varepsilon_\mathrm{CM}(\omega | a,b ) = 1 - \frac{\omega_p^2}{\omega_\tau^2} \left( 1 - ia\frac{1}{\omega_\tau \tau} + i b \frac{\omega \tau^{-1}}{\omega_\tau^2} \right)^{-1}.
\end{align}
Mermin's number conservation correction has been modified by $a \in [0,1]$ to smoothly turn off/on the correction. 
While, the momentum conservation correction has been modified by $b \in [-1,1]$ to either fully reverse momentum ($b=-1$) or fully conserve momentum ($b=1$) in a collision event.
The energy conservation corrections enter at order $k^2$, thus our completed Mermin and the Atwal-Ashcroft model predict the same expansion.
It is reasonable to make these conservation laws variable for a single species because we expect number conservation violating electron recombination events and momentum and energy conservation violating phonon scattering events.

\begin{figure}
    \centering
    \includegraphics[width=.5\textwidth]{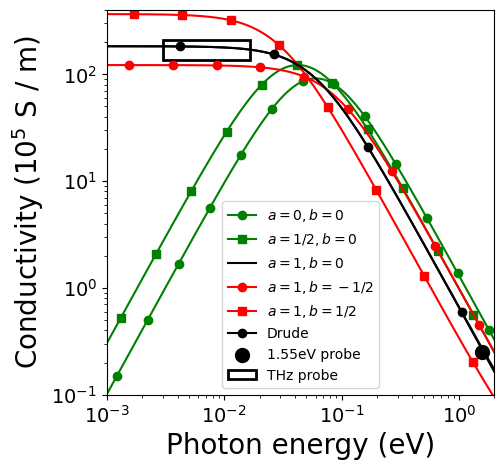}
    \caption{Chen et al.'s Drude conductivity fit (plotted as a solid black line) [Z. Chen, et al., Nature communications, 12.1, 1638, (2021)]. The completed Mermin conductivity model [defined from \eqref{eq_epsAAk0} and \eqref{eq_sigma-dielectric}] is plotted to demonstrate the effects of number conservation violation (green) and momentum preservation (red).}
    \label{fig_goldfits}
\end{figure}

Using \eqref{eq_sigma-dielectric} and \eqref{eq_epsAAk0} we produce a new semi-empirical conductivity model.
In Fig.~\ref{fig_goldfits}, we revisit Chen et al.'s Drude fit ($a=1$, $b=0$) to 300K gold data \cite{chen2021} and study the effects of parameters $a$ and $b$ on the shape of the conductivity model.
The central black line expresses their Drude fit. 
If particle number is not conserved ($a<1$), then, relative to the Drude model, the DC conductivity is suppressed and the optical conductivity is enhanced. 
Thus, the number conservation term primarily affects the model's slope at lower frequencies.
Whereas, if particle momentum is partially conserved ($b=1/2$), then, relative to the Drude model, the DC conductivity is enhanced and the optical conductivity is suppressed. Oppositely, if the collision reverses the momentum of a particle ($b=-1/2$), then the DC conductivity is suppressed and the optical conductivity is enhanced.
In total, the momentum conservation term primarily enhances or suppresses the DC conductivity, but does not change the slope in the low frequency region.

Lastly, our conductivity model can be compared to the primary alternative to the Drude model. 
The Drude-Smith (DS) conductivity model is defined
\begin{align}
    \sigma(\omega) = \frac{\sigma_0}{1 - i \omega \tau}\left( 1 + \frac{c}{1-i\omega \tau}\right).
\end{align}
Using \eqref{eq_sigma-dielectric}, the Drude-Smith (DS) dielectric function is given by
\begin{align}
    \varepsilon_\mathrm{DS}(\omega|c)  &= 1 - \frac{\omega_p^2}{\omega_\tau^2} \left( \frac{\omega_\tau+ i c / \tau }{\omega} \right). \label{eq_dielectric-DrudeSmith}
\end{align}



Comparing Smith's model \eqref{eq_dielectric-DrudeSmith} to our model \eqref{eq_epsAAk0}, we see that for $b=0$ (\textit{i.e.}, momentum conservation is turned off), then there is a direct relationship between our number conservation, $a$, and Smith's $c$.
This relation is expressed as,
\begin{align}\label{eq_DS-AA-relation}
    \left( \frac{\omega_\tau - i a/\tau}{\omega_\tau} \right)^{-1} = \left( \frac{\omega_\tau+ i c / \tau }{\omega} \right).
\end{align}
Notice, when number is conserved (\textit{i.e.}, $a=1$) and Smith's parameter is off (\textit{i.e.}, $c = 0$) Mermin's model and Smith's model are equivalent.
Further, when number is not conserved (\textit{i.e.}, $a=0$) and Smith parameter is on (\textit{i.e.}, $c = -1$) Mermin's model and Smith's model are equivalent.
This suggests Smith’s model breaks local number conservation to achieve DC conductivity suppression.


\section{Conclusions}
We have produced a new first-principles model of dynamical response for applications in XRTS diagnostics, experimental/theoretical conductivity estimates, stopping power estimates, and Bremsstrahlung emission estimates, which addresses sum rule violations and interpretation issues associated with existing models.  
We call this model the completed Mermin susceptibility; it is the first multi-species susceptibility that conserves number and momentum locally.
This result extends Selchow et al. \cite{Selchow1999} as well as Atwal and Ashcroft \cite{Atwal2002} from the single-species BGK to a multi-species BGK.
Affirming our expression, we recover the multi-species Mermin and RPA susceptibilities \cite{Selchow2001DSF} as limiting cases. Additionally, in the single species limit, our expression recovers the local field correction for momentum conservation derived in Morawetz and Fuhrmann \cite{Morawetz2000b}.

Our numerical and analytic results demonstrated our model satisfies both the f-sum rule and screening sum rule. In effect, the momentum conservation correction manifests as an infinitesimal plasmon peak, in the long wavelength limit, indicating that plasmons are the only energy loss mechanism when momentum is conserved. Additionally, we showed that in the single species limit, the momentum conservation correction enforces $\delta \j = n_0 \delta \u$. Comparatively Mermin's continuity equation does not assume that $\delta \j = n_0 \delta \u$, suggesting that f-sum rule violation occurs because of this.

We produced the two-component completed Mermin ion-ion DSF and applied the model to a plasma of deuterium and tritium at ICF hot-spot conditions. Comparing the completed Mermin DSF to the Mermin and RPA DSFs, we showed that conservation of momentum produces a peak near the ion plasma frequency not present in the Mermin and RPA models and that the light species approximation is in qualitative agreement with all mixed or unmixed deuterium or tritium plasmas. Using the light species approximation, we show that carbon contamination affects the Mermin DT DSF in a qualitatively different way than either the RPA and the completed Mermin DSF. This investigation establishes the model, but does not explore the impact of dynamical inter-species collision frequencies $\tau_{ij}(\omega)$ and effective ion-ion interaction potentials $v_{ij}(k)$, this will be the subject of future work. 

Finally, we applied our completed Mermin conductivity model to dynamical gold conductivity measurements. We showed that number conservation and moment conservation affect the dynamic conductivity differently. Our completed Mermin conductivity \eqref{eq_epsAAk0} can model back scattering with the parameters set to $a=1$ (locally number conserving) and $b \in [-1,0]$ (momentum reversed in the collision). This provides an interpretable alternative to the Drude-Smith conductivity model. However, when we vary the momentum reversal parameter $b$ the conductivity in both the DC and optical regimes is suppressed; this is unlike the behavior seen by varying the $c$ parameter in the Drude-Smith model. Therefore, we refute Smith's back-scattering interpretation. Additionally, by turning off ($a=0$) or on ($a=1$) local number conservation we produce the same functional form as seen in the Drude-Smith model at $c=-1$ or $c=0$ respectively. From this, we conclude that Smith's model violates number conservation to suppress the DC conductivity. We interpret the violation of local number conservation as the confining of free charge carriers. Thus, these results support Cocker et al.'s \cite{Cocker2017} charge carrier confinement interpretation.




\begin{acknowledgments}
Thomas Chuna would like to thank deceased MSU physicist F.D.C. Willard for lively discussions and insights into the foundational works of solid state physics. 
\end{acknowledgments}

\appendix

\section{Fourier transforming our kinetic equation \label{app_fourier}}
In this Appendix, we expand and linearize and Fourier Transform our multi-species BGK kinetic equation.
First, we list our chosen Fourier conventions.
We choose
\begin{align}
    \delta \fMG{i}(\r,\v, \omega) = \int dt \: e^{i \omega t} \delta \fMG{i}(\r,\v,t)
\end{align}
as the temporal convention; so that
\begin{align}\label{eq_Fourier-dfdt}
    &\int dt \: e^{i \omega t} \: \partial_t \:  \delta \fMG{i}(\r,\v,t) = - i \omega \: \delta \fMG{i}(\r,\v,\omega).
\end{align}
We choose
\begin{align}
    \delta \fMG{i}(\k,\v,t) = \int d\r \: e^{-i \k \cdot \r} \delta \fMG{i}(\r,\v,t)
\end{align}
as our spatial convention; so that
\begin{align}\label{eq_Fourier-dfdr}
    \int d\r \: e^{-i \k \cdot \r} \: \nabla_{\r} \delta \fMG{i}(\r,\v,t) =  i \k \:  \delta \fMG{i}(\k,\v,t).
\end{align}
This means that the usual expression for force
\begin{align}
    m \a^{(1)}_i(\r,t) = -\nabla_\mathbf{r} U^{(1)}_i(\r,t)
\end{align} 
leads to
\begin{align}\label{eq_Fourier-a}
    \a^{(1)}_i(\k,\omega) = - m^{-1} i \k \: U^{(1)}_i(\k, \omega).
\end{align}

We move to producing \eqref{eq_kinetic-helper1} from \eqref{eq_kinetic}. The BGK kinetic equation is given
\begin{align}
    \left( \partial_t + \v \cdot \nabla_\r  + \a^{(1)}_i \cdot \nabla_\v \right) f_i(\r,\v,t) = \sum_j Q_{ij}. 
\end{align}
We insert expansions \eqref{fiexpansion} and \eqref{Mijexpansion}, and then Fourier transform using \eqref{eq_Fourier-dfdt}, \eqref{eq_Fourier-dfdr}, and \eqref{eq_Fourier-a} to arrive at
\begin{align}\label{eq_kinetic-helper0}
    & \left( \v \cdot k - \omega \right) \lambda \delta \fMG{i}(\k,\v,\omega) \nonumber
    \\ & - \lambda \int d\k' d\omega' \: m_i^{-1} (\k - \k') \cdot \nabla_\v  \left( \fMG{i}(\v) \delta(k') \delta(\omega') + \delta \fMG{i}(\k,\v,\omega) \right) \: U^{(1)}_i(\k,\omega) \nonumber
    \\ & =  \sum_j \frac{i \lambda}{\tau_{ij}} \delta \fMG{i}(\k,\v,\omega) -  \sum_j \frac{i \lambda}{\tau_{ij}} \deltaM{ij}(\r,\v,t).
\end{align}
Notice that the term $\a^{(1)}_i \cdot \nabla_\v f_i(\r,\v,t)$ contains a multiplication of $U^{(1)}_i(\r,t)$ and $f_i(\r,\v,t)$, therefore it produces a convolution in Fourier space, which is the second term on the LHS of \eqref{eq_kinetic-helper0}.
We have replaced $f_i(\r,\v,t)$ with its expansion about global equilibrium $\fMG{i}(\v)$.
Since $\fMG{i}(\v)$ does not have position and time dependence, it is treated as a constant and its Fourier transform acquires Dirac delta functions.
From \eqref{eq_kinetic-helper0}, we drop $\delta \fML{i} \: U^{(1)}_i$ since it is second order and evaluate the convolution.
The simplified expression is
\begin{align}
    & \left( \v \cdot k - \omega \right) \lambda_i \delta \fMG{i}(\k,\v,\omega) - \lambda_i \: m_i^{-1} \k \cdot \nabla_\v \fMG{i}(\v)\: U^{(1)}_i(\k,\omega) \nonumber
    \\ & =  \sum_j \frac{i \lambda_i}{\tau_{ij}} \delta \fMG{i}(\k,\v,\omega) -  \sum_j \frac{i \lambda_i}{\tau_{ij}} \deltaM{ij}(\r,\v,t).
\end{align}
Canceling the fugacity factors and grouping the $\delta \fMG{i}(\k,\v,\omega)$ terms, we arrive at \eqref{eq_kinetic-helper1}.

\section{Evaluating the momentum conservation constraint \label{app_collisionoperatormoment}}
In this appendix, we show how to produce the momentum constraint, \eqref{eq_deltau}, from the first moment of the collision operator, \eqref{eq_multispeciesmomentumconservation}
\eqref{eq_multispeciesmomentumconservation}
\begin{align}
     \int d\v \: m_i \v Q_{ij} +  \int d\v \: m_j \v Q_{ji} = 0. 
\end{align}
From the definition of $Q_{ij}$, \eqref{eq_QMcBGK} and expansions \eqref{fiexpansion} and \eqref{Mijexpansion}, the expanded collision operator is
\begin{align} \label{eq_Qijexpansion}
    Q_{ij} \approx \frac{\lambda_i}{\tau_{ij}} \left( \deltaM{ij} - \delta \fMG{i} \right).
\end{align}
Inserting \eqref{eq_Qijexpansion} into the momentum constraint, canceling the fugacities, and grouping terms yields
\begin{align}
     \int d\v \: \v \left( m_i \frac{ \deltaM{ij} }{\tau_{ij}} +  m_j \frac{ \deltaM{ji} }{\tau_{ji}}\right) =  \int d\v \: \v \left( m_i \frac{\delta \fMG{i}}{\tau_{ij}} +  m_j \frac{\delta \fMG{j}}{\tau_{ji}} \right).
\end{align}
We insert the linear expansion for $\deltaM{ij}$, \eqref{dmijexpansion}, on the LHS and drop those terms which are odd in powers of $\v$ (Gaussian integrals evaluate to zero). The momentum constraint becomes
\begin{align}
     \frac{m_i}{\tau_{ij}} \int d\v  \: \v \frac{m_i^{-1}\k \cdot \nabla_\v f_{i,0}^M}{\v \cdot \k } \p_i \cdot \delta \u +  \frac{ m_j }{\tau_{ji}} \int d\v  \: \v \frac{m_j^{-1}\k \cdot \nabla_\v f_{j,0}^M}{\v \cdot \k } \p_j \cdot \delta \u =   \frac{m_i \delta \j_i}{\tau_{ij}} + \frac{m_j \delta \j_j}{\tau_{ji}}.
\end{align}
We used $\delta \j_i \equiv \int d\v \: \v \delta \fMG{i}$ to simplify the RHS. 
Finally, we evaluate the integrals on the LHS to produce \eqref{eq_deltau}. 

\section{Expanding partial fractions\label{app_fractionexpansion}}
In this appendix we integrate over velocity to produce the system’s linearized dynamical response given
\begin{align}
    \delta \fMG{i}(\k,\v,\omega) &= \frac{m_i^{-1} \: \k \cdot \nabla_\v \fMG{i}(\v)}{\v \cdot \k - \omega_{\tau_i} } \nonumber
    \\ & \quad \times \left( U^{(1)}_i + \frac{i}{\v \cdot \k } \sum_j \frac{1}{\tau_{ij}}  \left( \delta \mu_i +  \p_i \cdot \delta \u_{ij}
    + \left( \frac{p^2_i}{2m_i} - \mu_i \right)  \frac{\delta T_{ij}}{ T} \right) \right). 
\end{align}
The zeroth moment of this equation is expressed as
\begin{widetext}
\begin{align} \label{eq_kineticfAA-P}
    &\delta n_i =  U^{(1)}_i P \left[ \v \cdot \k \right] 
    + \frac{i}{\tau_i}  P \left[ 1 \right]  \delta \mu_i
    + \sum_j \frac{i}{\tau_{ij} } P \left[\p_i \cdot \delta \u\right]
    + \sum_j \frac{i}{\tau_{ij} } \left( P \left[ \frac{p_i^2}{2m_i} \right] - \mu_i P \left[ 1 \right] \right) \frac{\delta T_{ij}}{T}.
\end{align}
\end{widetext}
Here we have introduced the following linear functional $P$
\begin{align}\label{eq_P}
    P\left[ g \right] \equiv \int d\v \frac{m_i^{-1}\k \cdot \nabla_\v f_{i,0}^M}{\v \cdot \k \left(\v \cdot \k - \omega_{\tau_i} \right)} g.
\end{align}
These functionals are:
\begin{subequations} 
\begin{align}
    P\left[ p_i^{2n} (\v \cdot \k) \right] & = C^M_{i,2n}, \label{eq_Pp2vk}
    \\ P\left[ p_i^{2n} (\v \cdot \k - \omega_\tau) \right] &= B^M_{i,2n},
    \\ P\left[ \left(\frac{p_i^2}{2m_i} \right)^n \right] &= \frac{1}{\omega_\tau} \left( \frac{ C^M_{i,2n} }{(2m_i)^n} - \frac{B^M_{i,2n}}{(2m_i)^n} \right), \label{eq_P1}
    \\ P \left[\p_i \cdot \delta \u\right] & = \frac{m_i}{k^2} \: C^M_{i,0} \: \k \cdot \delta \u.
\end{align}
\end{subequations}
Notice that \eqref{eq_P1} at $n=0$ is $P\left[ 1 \right]$. 
Inserting these expressions for functional $P$ into \eqref{eq_kineticfAA-P} recovers \eqref{eq_deltan}.

\section{Implementing our model\label{app_specialfxns}}
In this appendix, we collect the information necessary to compute the multi-species completed Mermin susceptibility. 

\subsection{Expressing \texorpdfstring{$C^M_{i,2n}$}{} with dimensional quantities}
The completed Mermin susceptibility has been expressed in terms of $C^M_{i,2n}(\k,\omega)$. 
\begin{align}
    C^M_{i,2n}(\k,\omega) &\equiv \int d\v |\p|^{2n} \frac{m_i^{-1} \k \cdot \nabla_\v \fMG{i}}{\v \cdot \k - \omega}. \tag{\ref{eq_Cn}'}
\end{align}
Thus, our first step is create a dimensional version of this expression for $2n=0,2,4$.
In the absence of an external magnetic field, the x axis is arbitrarily chosen as the direction of the wave vector (\textit{i.e.} $\v \cdot \v = v_x^2$ and $\v \cdot \k = v_x k_x$ ).
We insert $\nabla_\v \fMG{i}$ and evaluate the $v_y$, $v_z$ integrals.
This recasts our function as
\begin{align}
    C^M_{i,2n} &=  -\frac{n_i}{T} (m_i T)^n \left( \frac{1}{2 \pi }\right)^{1/2} \int_{\mathcal{C}} d \tilde{v} \: \tilde{v}^{2n} \frac{\tilde{v} e^{-\tilde{v}^2/2} }{\tilde{v} -\sqrt{\frac{m_i}{T}} \frac{\omega_\tau}{k_x}}.
\end{align}
Here $\Tilde{v} = \sqrt{\frac{m_i}{T}} v_x$ is the dimensionless velocity.
We can express the term in the denominator in relation to the electron's dimensionless phase velocity
\begin{align}
    \sqrt{\frac{m_i}{T}} \frac{\omega_\tau}{k_x} &= \zeta_i \tilde{v}_{p}.
\end{align}
This equality follows from the definitions of the electron plasma frequency $\omega_{p,e}^2 \equiv 4 \pi n_e \: e^2 / m_e$, the electron Debye wavenumber $k_{D,e}^2 \equiv 4 \pi n_e \: e^2 /  T$, the mass fraction $\zeta_i \equiv \sqrt{m_i / m_e}$, and the dimensionless phase velocity $\tilde{v}_{p} \equiv \frac{\omega_\tau / \omega_{p,e} }{k_x / k_{D,e} }$.
We use the electron based normalization to make the multi-species calculations simpler. 
The final dimensionless expression of $C^M_{i,2n}$ is
\label{eq_dimensionlessChi}
\begin{subequations}
\label{eq_dimensionalC}
\begin{align}
    C^M_{i,2n}(k, \omega_\tau) &= -\frac{n_i}{T} (m_iT)^n F_{2n}(\zeta_i \tilde{v}_p),
    \\ B^M_{i,2n} &\equiv C^M_{i,0}(k,0), 
    \\ F_{2n}(z) &= \left( \frac{1}{2 \pi }\right)^{1/2} \int_{\mathcal{C}} d \tilde{v} \: \tilde{v}^{2n} \frac{\tilde{v}}{\tilde{v} - z}.
\end{align}
\end{subequations}
We note that $B^M_{i,0} = -\frac{n_i}{T}$, $B^M_{i,2} = -\frac{n_i}{T} m_iT$, and $B^M_{i,4} = -3 \frac{n_i}{T} (m_iT)^2$; this can be computed considering the $z \rightarrow 0$ limit.

\subsection{Expressing \texorpdfstring{$\chi$}{} with dimensionless quantities}
Our next step is to propagate these expressions for $C^M_{i,2n}$ and $B^M_{i,2n}$ into the completed Mermin susceptibility \eqref{eq_multispeciesCM}.
Consider the grouping
\begin{align}
    \varepsilon_a^* &=1 - v_{11}(k) C^M_{1,0} - \frac{i}{\omega_{\tau_1} \tau_1} \left( 1 - \frac{C^M_{1,0}}{B^M_{1,0}} \right) \nonumber
    \\ & \quad - i  \frac{ \omega}{k^2} \left( \frac{m_1}{n_{1,0} \tau_{11} }
    +  \frac{ \tau_{21} }{ \tau_{12} } \frac{m_1^2}{ m_1 n_{0,1} \tau_{21} + m_2 n_{0,2} \tau_{12}} \right) C^M_{1,0}.
\end{align}
Substituting $C^M_{i,2n}$ and $C^M_{i,2n}$ from \eqref{eq_dimensionalC} and $m_i/T = \zeta_i^2 k_{D,e}^2/ \omega_{D,e}^2$, we can recast the modified dielectric function as
\begin{align}
    \varepsilon_a^* &= 1 - v_{11}(k) C^M_{1,0} 
    - \frac{i}{\omega_{\tau_1} \tau_1} \left( 1 - F_0(\zeta_1 \tilde{v}_p) \right) \nonumber
    \\ & \quad + \left( i \frac{\zeta_1^2 \omega}{k^2 \tau_{11} } \frac{k_{D,e}^2}{\omega_{p,e}^2}
    + i \frac{ \omega }{k^2} \: \frac{ \tau_{21} }{ \tau_{12} } \frac{\psi^2_1 \zeta_1^4}{ \zeta^2_1 \psi^2_1 \tau_{21} + \zeta^2_2 \psi^2_2 \tau_{12}} \frac{k_{D,e}^2}{\omega_{p,e}^2} \right) F_0(\zeta_1 \tilde{v}_p).
\end{align}
We normalized our parameters to electronic scales $\tilde{k} = k / k_{D,e}$, $\tilde{\tau} = \tau \omega_{p,e}$, $\tilde{\omega} = \omega/\omega_{p,e}$, and $\tilde{\omega}_{\tau_1} = \omega_{\tau_1}/\omega_{p,e}$.
The result is
\begin{align}\label{eq_epsdimension-helper}
    \varepsilon_a^* &= 1 - v_{11}(k) C^M_{1,0} 
    - \frac{i}{ \tilde{\omega}_{\tau_1} \tilde{\tau}_1} \left( 1 - F_0 \right)
    + \left( i \frac{\zeta_1^2 \tilde{\omega}}{ \tilde{k}^2 \tilde{\tau}_{11} }
    + i \frac{ \tilde{\omega} }{\tilde{k}^2} \: \frac{ \tilde{\tau}_{21} }{ \tilde{\tau}_{12} } \frac{\psi^2_1 \zeta_1^4}{ \zeta^2_1 \psi^2_1 \tilde{\tau}_{21} + \zeta^2_2 \psi^2_2 \tilde{\tau}_{12}} \right) F_0,
\end{align}
where we have suppressed the arguments in $F_0( \zeta_1 \tilde{v}_p)$ for brevity.

Next, we non-dimensionalize  $v_{11}(k) C^M_{1,0}$, in \eqref{eq_epsdimension-helper}.
To do this, we assume that our potential $v_{ij}(k)$ is the Yukawa potential
\begin{align}
    v_{ij}(r) &=  \frac{Z_i Z_j e^2}{r} e^{- r / \lambda_s }.
\end{align}
We group terms into a dimensionless parameters $\tilde{r} =  k_{D,e} r $, $\tilde{\lambda}_s =\lambda_s k_{D,e}$ and Fourier transform over $\tilde{r}$ to arrive at 
\begin{align}
    v_{ij}(k) = \frac{T}{n_e} \left( \frac{Z_i Z_j}{\tilde{k}^2 + \tilde{\lambda}_s^{-2} } \right).
\end{align}
Notice that if we let $\lambda_s = k_{D,e}^{-1}$, as is the case in Thomas-Fermi theory, then $\tilde{\lambda}_s=1$. 
The factor of $T/n_\mathrm{e}$ will multiply $-n_i/T$ from the susceptibility and leave behind $n_i / n_\mathrm{e} = \psi_i^2$.
Therefore the product is expressed
\begin{align} \label{eq_dimensionalv}
    v_{ij}(k) C^M_{i,0} = -\psi_i^2 \left( \frac{Z_i Z_j}{\tilde{k}^2 + \tilde{\lambda}_s^{-2} } \right) F_0(\zeta_i \tilde{v}_p).
\end{align}
Inserting \eqref{eq_dimensionalv} into \eqref{eq_epsdimension-helper}, yields the final expression \begin{align}  \label{eq_dimensionaleps}
    \varepsilon_a^* &= 1 + \psi_i^2 \left( \frac{Z_1^2}{\tilde{k}^2 + \tilde{\lambda}_s^{-2} } \right) F_0(\zeta_1 \tilde{v}_p)
    - \frac{i}{ \tilde{\omega}_{\tau_1} \tilde{\tau}_1} \left( 1 - F_0(\zeta_1 \tilde{v}_p) \right)\nonumber
    \\ &\quad + \left( i \frac{\zeta_1^2 \tilde{\omega}}{ \tilde{k}^2 \tilde{\tau}_{11} }
    + i \frac{ \tilde{\omega} }{\tilde{k}^2} \: \frac{ \tilde{\tau}_{21} }{ \tilde{\tau}_{12} } \frac{\psi^2_1 \zeta_1^4}{ \zeta^2_1 \psi^2_1 \tilde{\tau}_{21} + \zeta^2_2 \psi^2_2 \tilde{\tau}_{12}} \right) F_0(\zeta_1 \tilde{v}_p).
\end{align}
The expression for $\varepsilon_b$ matches this one, but with indices $1$ and $2$ are switched. The only remaining expression in \eqref{eq_multispeciesCM} is the momentum conservation coupling term. Following the same steps as above yields
\begin{align} \label{eq_dimensionalcoupling}
    i \frac{\omega}{ k^2 } \frac{ m_1 m_2}{m_1 n_1 \tau_{21} + m_2 n_2 \tau_{12}} C^M_{1,0} = - i \frac{ \tilde{\omega} }{k^2} \frac{\zeta_1^2 \zeta_2^2}{ \zeta^2_1 \psi^2_1 \tilde{\tau}_{21} + \zeta^2_2 \psi^2_2 \tilde{\tau}_{12}} \psi_1^2 F_0(\zeta_1 \tilde{v}_p).
\end{align}
Together \eqref{eq_dimensionalv}, \eqref{eq_dimensionaleps}, and \eqref{eq_dimensionalcoupling} provide a dimensionless representation of the completed Mermin susceptibility \eqref{eq_multispeciesCM}.

\subsection{Representing \texorpdfstring{$F_{2n}$}{} with special functions}
To evaluate \eqref{eq_dimensionalv}, \eqref{eq_dimensionaleps}, and \eqref{eq_dimensionalcoupling}, we need numerical implementations of $F_0(z)$.
Ichimaru Ch.4, expresses $F_0(z)$ in terms of the special function $W(Z)$ \cite{Ichimaru1991}. We extended Ichimaru's procedure to calculate $F_2(z)$ and $F_4(z)$.
The resulting expressions are
\begin{subequations}
\begin{align}
    &F_0(z) = 1 + z \left( i\sqrt{\frac{\pi}{2}} \text{WofZ}(z / \sqrt{2}) \right) ,
    \\& F_2(z) = 1 + z^2 + z^3 \left( i\sqrt{\frac{\pi}{2}} \text{WofZ}(z / \sqrt{2}) \right),
    \\& F_4(z) = 3 + z^2 + z^4 + z^5 \left( i\sqrt{\frac{\pi}{2}} \text{WofZ}(z / \sqrt{2}) \right),
    \\& \text{WofZ}(\frac{z}{\sqrt{2}}) \equiv i\sqrt{\frac{2}{\pi}}e^{- z^2/2} \int_0^z d\tilde{v} e^{\tilde{v}^2/2} + e^{-z^2 /2}.
\end{align}
\end{subequations}
We denote $W(z)$ as WofZ$(z)$ to align with scipy's notation \cite{2020SciPy-NMeth}.
The only remaining unknowns are the parameters that describe our plasma system: $\psi_i$, $\zeta_i$, $Z_i$, $\tilde{\lambda}_s$, $\nu_{ij}$.

\section{Expanding the dielectric function at long wavelengths \label{app_expansions}}
Here, we compute the long-wavelength expansions presented in \eqref{eq_longwavelengthexpansion}.
For the OCP, \eqref{eq_dielectric} informs us that
\begin{align} \label{eq_ss-dielectric}
    \frac{1}{\varepsilon (k,\omega)} = 1 + v_{11}(k) \chi_{11}(k,\omega).
\end{align}
Here $\chi_{11}$ is the mean field corrected susceptibility. 
We rewrite \eqref{eq_ss-dielectric} in terms of susceptibilities without mean field corrections and since there is only one species, we will suppress the indices \textit{i.e.}, $\chi_{11} = \chi_0/ ( 1 - v(k) \chi_0)$. The result is
\begin{align}\label{eq_dielectric-helper0}
    \frac{1}{\varepsilon(k,\omega)} = \frac{1}{ 1 - v(k) \chi_0 }.
\end{align}
By conducting a long wave expansion on \eqref{eq_dielectric-helper0}, we can compute \eqref{eq_longwavelengthexpansion}.

Starting from Mermin \eqref{eq_singlespeciesM}, completed Mermin \eqref{eq_singlespeciesCM}, and Atwal-Ashcroft \cite{Atwal2002}, we follow the same steps as Appendix \ref{app_specialfxns} to produce non-dimensional representations of the single species susceptibilities \textit{without} mean field correction.
These expressions are
\begin{subequations}
\label{eq_singlespeciessusceptibilties}
\begin{align}
    \chi^M_0 &\equiv \frac{n}{T} \frac{- F_0}
    { \frac{\tilde{\omega}}{\tilde{\omega}_\tau}  + \frac{i}{\tildetwc} F_0},
    \\ \chi^{CM}_0 &\equiv \frac{n}{T} \frac{- F_0 }
    { \frac{\tilde{\omega}}{\tilde{\omega}_\tau}  + i \left( \frac{1}{\tildetwc} + \frac{ \zeta^2 \tilde{\omega}}{ \tilde{k}^2 \tilde{\tau}} \right) F_0 }, 
    \\ \chi^{AA}_0 &\equiv \frac{n}{T}
    \frac{ - F_0 + \frac{i}{2 \tildetw} \left(F_2 F_2 - F_0 F_4 \right) }
    {\frac{\tilde{\omega}}{\tilde{\omega}_\tau} + \frac{i \zeta^2 \tilde{\omega} }{k^2 \tilde{\tau}} F_0 + (\frac{1}{2 \tilde{\omega} \tilde{\omega}_\tau \tilde{\tau}^2 } + \frac{\zeta^2}{2 \tilde{\tau}^2 \tilde{k}^2 })(F_2 F_2 - F_0 F_4 ) + \frac{i}{ 2 \tildetwc} \left( 3 F_0 - 2F_2 + F_4 \right) },
\end{align}
\end{subequations}
where we have normalized our parameters to electronic scales: $\zeta^2 = m/m_e$, $\tilde{k} = k / k_{D,e}$, $\tilde{\tau} = \tau \omega_{p,e}$, $\tilde{\omega} = \omega/\omega_{p,e}$, and $\tilde{\omega}_{\tau} = \omega_{\tau}/\omega_{p,e}$.

As shown in Appendix \ref{app_specialfxns}, $F_0(z)$, $F_2(z)$, and $F_4(z)$ can be expressed in terms of the WofZ function, which has known $z \rightarrow \infty$ expansions.
Expanding $F_0(z)$, $F_2(z)$, and $F_4(z)$ at large $z$ yields the following analytic expressions:
\begin{subequations} 
\label{eq_Fnexpansion}
\begin{align}
    \underset{z \rightarrow \infty}{\lim} F_0(z)  &= -1/z^2 + \mathcal{O}[z^{-4}],
    \\ \underset{z \rightarrow \infty}{\lim} F_2(z)  &= -3/z^2 + \mathcal{O}[z^{-4}],
    \\ \underset{z \rightarrow \infty}{\lim} F_4(z)  &= -15/z^2 + \mathcal{O}[z^{-4}].
\end{align}
\end{subequations}
To construct \eqref{eq_longwavelengthexpansion}, we first substitute \eqref{eq_Fnexpansion} into the susceptibilities \eqref{eq_singlespeciessusceptibilties} and then substitute \eqref{eq_singlespeciessusceptibilties} into \eqref{eq_dielectric-helper0}. Next, we make the denominator real and collect the imaginary parts. Finally, we simplify the expressions using $n \rightarrow n_e$ and $\zeta \rightarrow 1$; this leaves behind only factors of $\tilde{\omega}$, $\tilde{\tau}$, and $\tilde{k}$.

\bibliography{references}

\end{document}